\documentclass[epj,nopacs]{svjour}

\usepackage{graphicx}
\usepackage{latexsym}
\usepackage{amsmath}
\usepackage{amssymb}
\usepackage{amsfonts}
\usepackage{bm}
\usepackage{color}

\newcommand{\Fcal}{\ensuremath{{\cal F}}}

\newcommand{\Ncal}{\ensuremath{{\cal N}}}
\newcommand{\Ocal}{\ensuremath{{\cal O}}}
\newcommand{\Pcal}{\ensuremath{{\cal P}}}

\newcommand{\Ucal}{\ensuremath{{\cal U}}}

\newcommand{\Eop}{\ensuremath{\mathbf{E}}}

\newcommand{\Sop}{\ensuremath{\mathbf{S}}}
\newcommand{\Vop}{\ensuremath{\mathbf{V}}}

\newcommand{\ddiff}{\ensuremath{\mathrm{d}}}

\newcommand{\bfzero}{\ensuremath{{\bf 0}}}

\newcommand{\qvec}{\ensuremath{{\bf q}}}
\newcommand{\rvec}{\ensuremath{{\bf r}}}

\newcommand{\qx}{q_x}
\newcommand{\qy}{q_y}
\newcommand{\rx}{r_x}
\newcommand{\ry}{r_y}

\newcommand{\la}{\left<}
\newcommand{\ra}{\right>}

\newcommand{\kB}{k_\mathrm{B}}

\newcommand{\dVcell}{\delta V}
\newcommand{\agrid}{a_{\mathrm{grid}}}
\newcommand{\ngrid}{n_{\mathrm{grid}}}

\newcommand{\ar}{a_{\rvec}}

\newcommand{\br}{b_{\rvec}}

\newcommand{\ccr}{c_{\rvec}}

\newcommand{\Er}{E_{\rvec}}
\newcommand{\fq}{f_{\qvec}}
\newcommand{\fr}{f_{\rvec}}
\newcommand{\gq}{g_{\qvec}}
\newcommand{\gr}{g_{\rvec}}
\newcommand{\mq}{m_{\qvec}}
\newcommand{\mr}{m_{\rvec}}

\newcommand{\kr}{k_{\rvec}}
\newcommand{\vq}{v_{\qvec}}
\newcommand{\vr}{v_{\rvec}}

\newcommand{\xr}{x_{\rvec}}

\newcommand{\yr}{y_{\rvec}}

\newcommand{\tincr}{\delta \tau}
\newcommand{\tsamp}{\Delta \tau}

\newcommand{\taubasin}{\tau_{\mathrm{b}}}
\newcommand{\tspacer}{\tau_{\mathrm{spacer}}}

\newcommand{\xbf}{\ensuremath{\mathbf{x}}}

\newcommand{\Nt}{\ensuremath{N_\mathrm{t}}}
\newcommand{\Nr}{\ensuremath{N_{\rvec}}}
\newcommand{\Nc}{\ensuremath{N_\mathrm{c}}}
\newcommand{\Nk}{\ensuremath{N_\mathrm{k}}}
\newcommand{\Nl}{\ensuremath{N_l}}
 
\newcommand{\Om}{\ensuremath{m}}
\newcommand{\Ov}{\ensuremath{v}}

\newcommand{\Ockq}{\ensuremath{\Ocal_{ck\qvec}}}
\newcommand{\Ockr}{\ensuremath{\Ocal_{ck\rvec}}}
\newcommand{\Ocq}{\ensuremath{\Ocal_{c\qvec}}}
\newcommand{\Ocr}{\ensuremath{\Ocal_{c\rvec}}}

\newcommand{\dOtot}{\ensuremath{\delta \Ocal^2_{\mathrm{tot}}}}
\newcommand{\dOint}{\ensuremath{\delta \Ocal^2_{\mathrm{int}}}}
\newcommand{\dOext}{\ensuremath{\delta \Ocal^2_{\mathrm{ext}}}}
\newcommand{\sOtot}{\ensuremath{\delta \Ocal_{\mathrm{tot}}}}
\newcommand{\sOint}{\ensuremath{\delta \Ocal_{\mathrm{int}}}}
\newcommand{\sOext}{\ensuremath{\delta \Ocal_{\mathrm{ext}}}}

\newcommand{\svtot}{\ensuremath{\delta v_{\mathrm{tot}}}}
\newcommand{\dvint}{\ensuremath{\delta v^2_{\mathrm{int}}}}
\newcommand{\svint}{\ensuremath{\delta v_{\mathrm{int}}}}
\newcommand{\dvext}{\ensuremath{\delta v^2_{\mathrm{ext}}}}
\newcommand{\svext}{\ensuremath{\delta v_{\mathrm{ext}}}}
\newcommand{\dvgauss}{\ensuremath{\delta v^2_{\mathrm{G}}}}
\newcommand{\svgauss}{\ensuremath{\delta v_{\mathrm{G}}}}

\newcommand{\Khat}{\ensuremath{K}}
\newcommand{\Kimp}{\ensuremath{C_{\mathrm{imp}}}}
\newcommand{\Ctot}{\ensuremath{C_{\mathrm{tot}}}}
\newcommand{\Cint}{\ensuremath{C_{\mathrm{int}}}}
\newcommand{\Cext}{\ensuremath{C_{\mathrm{ext}}}}

\newcommand{\xiind}{\ensuremath{\xi_{\mathrm{ind}}}}
\newcommand{\xistar}{\ensuremath{\xi_{\star}}}

\newcommand{\Dnonerg}{\Delta_{\mathrm{ne}}^2}
\newcommand{\Snonerg}{\Delta_{\mathrm{ne}}}
\newcommand{\Tnonerg}{\tau_{\mathrm{ne}}}

\newcommand{\rnonerg}{r_{\mathrm{ne}}}

\begin{document}

\title{Spatial correlation functions for non-ergodic stochastic processes of macroscopic systems}
\titlerunning{Spatial correlation functions for non-ergodic stochastic processes}

\author{J.P.~Wittmer\thanks{joachim.wittmer@ics-cnrs.unistra.fr}
\and A.N. Semenov
\and J. Baschnagel
}
\institute{Institut Charles Sadron, Universit\'e de Strasbourg \& CNRS, 23 rue du Loess, 67034 Strasbourg Cedex, France}
\date{Received: date / Revised version: date}

\abstract{Focusing on non-ergodic macroscopic systems 
we reconsider the variances $\delta \Ocal^2$ of time averages $\Ocal[\xbf]$ 
of time-series $\xbf$. 
The total variance $\dOtot = \dOint + \dOext$ (direct average over all time-series) 
is known to be the sum of 
an internal variance $\dOint$ (fluctuations within the meta-basins)
and an external variance $\dOext$ (fluctuations between meta-basins).
It is shown that whenever $\Ocal[\xbf]$ can be expressed
as a volume average of a local field $\Ocal_{\rvec}$
the three variances can be written 
as volume averages of correlation functions $\Ctot(\rvec)$, $\Cint(\rvec)$
and $\Cext(\rvec)$ with $\Ctot(\rvec) = \Cint(\rvec) + \Cext(\rvec)$.
The dependences of the $\delta \Ocal^2$ on the sampling time $\tsamp$ and the system volume $V$
can thus be traced back to $\Cint(\rvec)$ and $\Cext(\rvec)$.
Various relations are illustrated using lattice spring models with 
spatially correlated spring constants.
}
\date{\today}
\maketitle

\section{Introduction}
\label{intro}

\subsection{Background}
\label{intro_background}

Let us consider a stochastic dynamical variable $x(\tau)$,
like certain density or stress fields averaged over the system volume $V$,
characterizing a large physical system as a function of (continuous) time $\tau$. 
Extending recent work on stationary stochastic processes in non-ergodic macroscopic systems 
\cite{lyuda19a,spmP1,spmP2,spmP3} we investigate here quite generally the variances 
$\delta \Ocal^2(\tsamp,V)$ of observables $\Ocal[\xbf]$ of time-series $\xbf$. 
As further specified in Sec.~\ref{spf}, a time-series $\xbf$ stands for an ensemble 
of discrete data entries $x_t$ sampled over a ``sampling time" $\tsamp$ and $\Ocal[\xbf]$ 
for a time-averaged moment over the data entries $x_t$.
While for ergodic systems independently created configurations $c$
are able in principle given enough time to explore the complete (generalized) phase space, 
for strictly non-ergodic systems they are permanently trapped in meta-basins 
\cite{GoetzeBook,Heuer08}. 
The different time-series $k$ of the same independent configuration $c$ are then correlated 
being all confined to the same basin even if separated by arbitrarily long spacer
(tempering) time intervals \cite{spmP2}. 
A time-series $\xbf_{ck}$ must now be characterized by {\em two} indices $c$ and $k$
and it becomes crucial in which order $c$- and $k$-averages are taken.
This implies that the commonly used total variance \cite{spmP2} 
\begin{equation}
\dOtot(\tsamp,V) = \dOint(\tsamp,V) + \dOext(\tsamp,V)
\label{eq_intro_dOtot}
\end{equation}
becomes the sum of two independent terms:
an internal variance $\dOint$, measuring the typical fluctuations within each meta-basins,
and an external variance $\dOext$, comparing the different meta-basins.
Importantly, $\sOint$ and $\sOext$ depend differently on $\tsamp$ and $V$.
For $\tsamp$ larger than the typical relaxation time $\taubasin$ of the 
meta-basins,
$\sOint(\tsamp,V)$ decays as $\sqrt{\taubasin/\tsamp}$ while $\sOext(\tsamp,V)$
becomes a $\tsamp$-in\-de\-pen\-dent constant. This large-$\tsamp$ limit
\begin{equation}
\Snonerg(V) \equiv \lim_{\tsamp\to \infty} \sOext(\tsamp,V)
\label{eq_intro_Snonerg}
\end{equation}
is our definition of the  
``non-ergodicity parameter" \cite{spmP1,spmP2,spmP3},\footnote{Other 
definitions of non-ergodicity parameters 
may be found in the literature \cite{GoetzeBook}. Our definition
does not rely on the properties of a specific model or a theoretical assumption.
It can be made system-size independent by multiplying it with the
appropriate non-universal volume dependence $V^{\gamma}$.}
an important order parameter vanishing for ergodic stochastic processes 
but remaining positive definite for non-ergodic systems \cite{spmP2}.
 
\subsection{Motivations}
\label{intro_motivation}

For macroscopic systems {\em without} long-range spatial correlations it is not difficult
to predict the system-size scaling of $\Snonerg(V)$ \cite{spmP2}.
Quite generally, this leads to a power law $\Snonerg(V) \simeq 1/V^{\gamma}$
where the exponent $\gamma$ naturally depends on the considered observable $\Ocal[\xbf]$.
Deviations from this exponent suggest long-range correlations.
Such deviations have, e.g., been observed for the non-ergodicity parameter $\Snonerg(V)$
associated to the elastic shear shear modulus \cite{spmP1,Procaccia16}
obtained by means of the stress-fluctuation formalism \cite{Lutsko88,Lutsko89,Barrat06,WXP13}
or the (closely related) variance of the shear stresses \cite{lyuda19a,spmP1,spmP2,spmP3} 
in viscoelastic and/or glass-forming colloidal systems.
Unfortunately, it becomes numerically ra\-pid\-ly demanding to
precisely obtain $\Snonerg(V)$ for increasingly large systems
and, quite generally, it gets impossible to characterize the spatial correlations 
just by measuring the $V$-dependence of macroscopic properties such as $\Snonerg(V)$.
It is thus crucial to directly measure the correlations \cite{Lemaitre14,Lemaitre15}
and to do this {\em consistently} with the non-ergodicity of the systems. 

\subsection{New key results}
\label{intro_novel}

We assume in the present work that the macroscopic observable $\Ocal[\xbf]$ 
can be written as a linear superposition $\Ocal[\xbf] = \Eop^{\rvec} \Ocal_{\rvec}$
of an associated local field $\Ocal_{\rvec}$.
(Using the notation introduced in Sec.~\ref{pre_notations},
$\Eop^{\rvec}$ denotes here a spatial average over microcells 
at a position $\rvec$ of the system.)
One main point of the present study is to show
that it is then both possible and useful 
to write the three different variances as volume averages
\begin{eqnarray}
\dOtot(\tsamp,V) & = & \Eop^{\rvec} \Ctot(\rvec,\tsamp,V) \label{eq_intro_Ctotr2dOtot} \\
\dOint(\tsamp,V) & = & \Eop^{\rvec} \Cint(\rvec,\tsamp,V) \label{eq_intro_Cintr2dOint} \\ 
\dOext(\tsamp,V) & = & \Eop^{\rvec} \Cext(\rvec,\tsamp,V) \label{eq_intro_Cextr2dOext} 
\end{eqnarray}
over the corresponding spatial correlation functions $\Ctot$, $\Cint$ and $\Cext$
which are properly defined below in Sec.~\ref{corr_inexto} and Appendix~\ref{app_corr}, 
Eqs.~(\ref{eq_Ctotr_def}-\ref{eq_Cintr_def}).
Moreover, this can be done in such a way that 
\begin{equation}
\Ctot(\rvec,\tsamp,V) = \Cint(\rvec,\tsamp,V) + \Cext(\rvec,\tsamp,V)
\label{eq_intro_Cinexto}
\end{equation}
holds in analogy of Eq.~(\ref{eq_intro_dOtot}).
This makes it possible to trace back the $\tsamp$- and $V$-dependences of the three different 
macroscopic variances to the {\em two} correlation functions $\Cint$ and $\Cext$.
Various relations will be illustrated by means of a simple ``Lattice Spring Model" (LSM) 
characterized by two quenched and spatially correlated lattice fields.

\subsection{Outline}
\label{intro_outline}

We begin by addressing in Sec.~\ref{pre} some technicalities
such as useful conventions (Sec.~\ref{pre_notations} and Sec.~\ref{pre_grid}),
the description of the LSM (Sec.~\ref{pre_LSM} and Sec.~\ref{pre_CarCbr})
and the determination and use of spatial correlation functions 
(Sec.~\ref{pre_Cr} and Sec.~\ref{pre_Cr2OV}).
We construct then in Sec.~\ref{spf} from the instantaneous
stoch\-as\-tic processes $x_t$ and fields $x_{t\rvec}$ (Sec.~\ref{spf_x}) 
the time-averaged observables $\Ocal[\xbf]$ and fields $\Ocal_{\rvec}$ (Sec.~\ref{spf_Ox})
Summarizing Refs.~\cite{spmP1,spmP2,spmP3} we remind in Sec.~\ref{glob_reminder} 
various features of the three variances $\delta \Ocal^2$. 
The corresponding $\tsamp$- and $V$-dependences are illustrated in, respectively, 
Sec.~\ref{glob_tsamp} and Sec.~\ref{glob_Vol} using numerical results obtained by 
means of LSM simulations.
We turn in Sec.~\ref{corr} to the spatial correlation functions.
Examples from the LSM simulations are discussed in Sec.~\ref{corr_examples}.
We conclude the paper in Sec.~\ref{conc} where we shall finally
hint on results of preliminary related work investigating the internal and external 
spatial correlation functions of time-averaged fields obtained from instantaneous 
stress fields in amorphous colloidal glasses.
The derivation of Eqs.~(\ref{eq_intro_Ctotr2dOtot}-\ref{eq_intro_Cinexto})
is presented in Appendix~\ref{app_corr}. 
The general scaling of the internal correlation function 
$\Cint[v](\rvec,\tsamp,V)$ of the important local variance field $\Ocal_{\rvec}=v_{\rvec}$
(as defined in Sec.~\ref{spf_Ox})
is discussed in detail in Appendix~\ref{app_Cint}.

\section{Conventions and technicalities}
\label{pre}

\subsection{Notations}
\label{pre_notations}

We use the same compact operator notations as in Refs.~\cite{spmP2,spmP3}.
The arithmetic $l$-average operator 
\begin{equation}
\Eop^l \Ocal_{lmn\ldots}  
\equiv \frac{1}{\Nl} \sum_{l=1}^{\Nl} \Ocal_{lmn\ldots} \label{eq_Eop_def}
\end{equation}
takes a property $\Ocal_{lmn\ldots}$ depending possibly on several indices $l,m,\ldots$
and projects out the specified index $l$, 
i.e. the $l$-average $\Ocal_{mn\ldots}(\Nl) \equiv \Eop^l \Ocal_{lmn\ldots}$ 
does not depend on $l$, but as marked by the argument may depend on the upper bound $\Nl$.
The $l$-variance operator $\Vop^l$ is {\em defined} by
\begin{eqnarray}
\Vop^l \Ocal_{lmn\ldots} 
& \equiv & \Eop^l \Ocal_{lmn\ldots}^2 - \left(\Eop^l\Ocal_{lmn\ldots}\right)^2 \nonumber \\
& \equiv & \frac{1}{\Nl} \sum_{l=1}^{\Nl} \left( \Ocal_{lmn\ldots}- \Ocal_{mn\ldots}\right)^2.
\label{eq_Vop_def} 
\end{eqnarray}
Note that the ``empirical variance" 
$\delta \Ocal_{mn\ldots}^2(\Nl) \equiv \Vop^l \Ocal_{lmn\ldots}$ vanishes
\begin{equation}
\delta \Ocal_{mn\ldots}^2(\Nl) \to 0 \mbox{ for } \Nl \to 1.
\label{eq_Vop_Nlsmall}
\end{equation}
In many cases $\Ocal_{mn\ldots}(\Nl)$ and 
$\delta \Ocal_{mn\ldots}(\Nl)$ converge for large $\Nl$ 
or become stationary for a large $\Nl$-window of 
the experimentally or numerically accessible $\Nl$-range.
To simplify notations we often denote this limit by 
$\Ocal_{mn\ldots}$ and $\delta \Ocal_{mn\ldots}$ {\em without} the argument $\Nl$.
As discussed in Ref.~\cite{spmP2}, 
we have defined the empirical variance as an uncorrected (biased) sample variance 
without the usual Bessel correction \cite{LandauBinderBook}, 
i.e. we normalize in Eq.~(\ref{eq_Vop_def}) with $\Nl$ and not with $\Nl-1$. 
This implies 
\begin{equation}
\delta \Ocal_{mn\ldots}^2(\Nl) \simeq \left(1 - \frac{1}{\Nl} \right) \delta \Ocal_{mn\ldots}^2
\label{eq_Vop_Nl} 
\end{equation}
for variances obtained with finite $\Nl$.
This relation is used below to extrapolate finite-$\Nl$ observables to $\Nl \to \infty$.

\subsection{Periodic grid of microcells}
\label{pre_grid}

\begin{figure}[t]
\centerline{\resizebox{.80\columnwidth}{!}{\includegraphics*{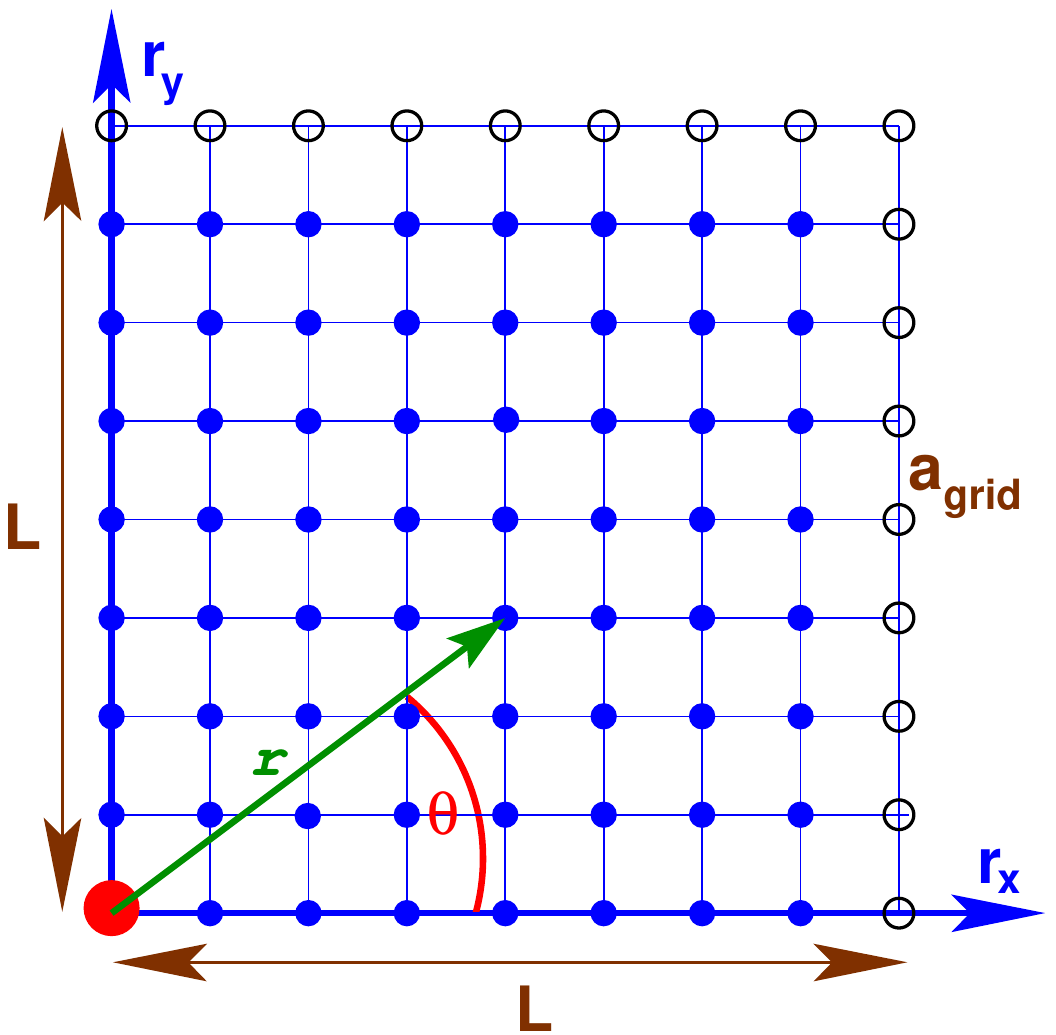}}}
\caption{Sketch of a two-dimensional ($d=2$) 
square lattice with $\ngrid=L/\agrid=2^3$. 
The filled circles indicate microcells of the principal box,
the open circles some periodic images.
The spatial position $\rvec$ of a microcell is either given by the $\rx$- and 
$\ry$-coordinates (in the principle box) or by the distance $r=|\rvec|$ 
from the origin (large circle) and the angle $\theta$.
}
\label{fig_grid_sketch}
\end{figure}

We shall illustrate below various properties by means of (real and discrete) 
fields $\fr$ ($\rvec$ labeling the microcell position) corresponding to $\Nr$ microcells 
on regular grids in $d=2$ dimensions as sketched in Fig.~\ref{fig_grid_sketch}.
For simplicity we assume square periodic lattices of linear dimension $L=\ngrid \agrid$, 
i.e. of ($d$-dimensional) volume 
$V=L^d = \Nr \dVcell$ with $\dVcell=\agrid^d$ being the microcell volume. 
To characterize spatial correlations (cf. Sec.~\ref{pre_Cr})
it is convenient \cite{numrec} to focus not on $\fr$
but on its discrete Fourier transform
\begin{equation}
\fq = \Fcal\{\fr\} = \Eop^{\rvec} \fr \exp(i \qvec \cdot \rvec)
\label{eq_fq}
\end{equation}
with $\Eop^{\rvec}$ being the average over all $\Nr$ microcells,
using the notation Eq.~(\ref{eq_Eop_def}), and $\qvec$ the discrete wavevectors
(being commensurate with the periodic grid).\footnote{The
inverse Fourier transform is $f_{\rvec}= \sum_{\qvec} f_{\qvec} \exp(-i \qvec \cdot \rvec)$.}
Due to our Fourier transform convention the sum rule $f_{\qvec=0} = \Eop^{\rvec} \fr$ holds.
We denote by $f$ the field irrespective of its representation and specific values
while $\fr$ refers to the instantaneous field in real space
and $\fq$ to the corresponding discrete field in reciprocal space.
Fast Fourier Transforms (FFT) \cite{numrec} are naturally used
for the efficient transformation between real and reciprocal space and
it is thus convenient to set $\ngrid$ to be an integer power of $2$.
Moreover, we set $\agrid=1$, i.e. $L=\ngrid$ and $V=\Nr$.

\begin{table*}[t]
\begin{center}
\begin{tabular}{|c|l|l|l|l|}
\hline
LSM & description             & $C[a](\rvec)$                                          & $C[b](\rvec)$                  & $J$  \\ \hline
 A    & uncorrelated sites      & $\ar \in \Ucal[-0.1,0.1]$                                  & $\br \in \Ucal[0.1,1.9]$            & 0  \\
 B    & exponential decay       & $C[a](\rvec) \approx c_0 \exp(-r/\xi)$                     & $\ccr=\ar$,$\br=1+\ccr^2$      & 0, 0.1, 1  \\
 C    & power law decay         & $C[a](\rvec) \approx c_0/r^{\alpha}$ with $\alpha=1,2,3$ & $\ccr=\ar$,$\br=(1+0.1\ccr)^2$ & 0  \\
D    & anisotropic decay       & $C[a](\qvec)= 8\pi c_0(\qx \qy/q^2)^2/\Nr$                     
      & $\ccr=\ar$,$\br=(1+0.1\ccr)^2$ 
      &  $-1, 0, \ldots, 10$ \\ \hline
\end{tabular}
\caption[]{LSM variants studied with the third column 
indicating the imposed $C[a](\rvec)$ and the forth $C[b](\rvec)$.
The uncorrelated random fields $\ar$ and $\br$ of LSM-A are taken from the given 
uniform distributions $\Ucal[\ldots]$.
In all other cases $\br$ is computed using the indicated relation from an auxiliary field $\ccr$.
Naturally, the inverse spring constant $\br=1/\kr$ is always positive.
The coupling parameter $J$ (fifth column) for springs of neighboring grid sites is switched off 
but for LSM-B and LSM-D.
All correlations are isotropic for LSM-B and LSM-C while they are anisotropic for LSM-D. 
\label{tab_model}}
\end{center}
\end{table*}

\subsection{Lattice spring model}
\label{pre_LSM}

We present below MC simulations of a ``Lattice Spring Model" (LSM) with $\xr$ being the 
linear length of the ideal springs and $\ar$ and $\br = 1/\kr > 0$ two quenched fields 
imposing, respectively, the average length of a spring and its variance.
In addition, neighboring springs may be coupled by tuning a ``coupling parameter" $J$.
The energy $\Er$ of a microcell at $\rvec$ is thus given by
\begin{equation}
\Er = \frac{\kr}{2} \left( \xr - \ar \right)^2 
	+ \frac{J}{2} \sum_{\rvec'} (x_{\rvec'}-\xr)^2
\label{eq_Erspring}
\end{equation}
where the sum over $\rvec'$ runs over the $2d$ nearest-neighbors 
of $\rvec$ on the periodic grid. 
In the limit where the interactions between springs are switched off ($J=0$) or are small, 
this implies the thermal averages $\la \xr \ra = \ar$ and $\la \delta \xr^2 \ra = \br T > 0$
with $T$ being the temperature and setting Boltzmann's constant $\kB$ to unity.
We impose $T=1$ in all presented simulations.
A summary of the studied model variants is given in Table~\ref{tab_model}.
How spatially correlated fields $f=a$ and $f=b$ are generated
is explained in Sec.~\ref{pre_CarCbr}.
Using these fields we perform Metropolis MC simulations with local moves
\cite{LandauBinderBook,AllenTildesleyBook}.
Results are recorded in time intervals $\tincr=10$ measured in MC steps.

\subsection{Spatial correlation functions}
\label{pre_Cr}

In this work we shall impose or sample auto-corre\-la\-tion functions 
$C[f](\rvec) = \langle \Khat[\fr](\rvec) \rangle - \la \Eop^{\rvec} \fr \ra^2$ 
of various fields $f$.\footnote{We
note $f$ for the functional argument of the averaged correlation function $C$
and $\fr$ for the functional argument of the non-averaged correlation function $\Khat$.}
$\la \ldots \ra$ stands here for some general average (to be specified below),
$\rvec$ for any site (microcell) of the principal simulation box and 
\begin{equation}
\Khat[\fr](\rvec) \equiv \Eop^{\rvec'} f_{\rvec+\rvec'} f_{\rvec'}
\label{eq_Khat_def}
\end{equation}
for the non-averaged correlation function of one given field $\fr$.
All correlation functions are even and periodic (Fig.~\ref{fig_grid_sketch}).
Periodicity is most readily implemented in reciprocal space 
using the Wiener-Khinchin theorem (WKT) \cite{numrec}
\begin{equation}
\Khat[\fq](\qvec) \equiv \Fcal\{\Khat[\fr](\rvec)\} = |\fq|^2 = \fq f_{-\qvec}
\label{eq_WKT}
\end{equation}
for $\fq = \Fcal\{\fr\}$.
The Fourier transformed auto-correlation functions are thus
real and positive for all wavevectors $\qvec$. 
 
All (averaged) correlation functions $C[f](\rvec)$ or $C[f](\qvec)$ 
considered here have in addition $x \leftrightarrow y$-symmetry, but are not necessarily radial 
symmetric (isotropic) \cite{spmP3}. 
Instead of the $d$-dimensional fields $C[f](\rvec)$ we present below the weighted projections 
\begin{equation}
C[f,p](r=|\rvec|) \equiv \la C[f](r,\theta) \cos(2 p \theta) \ra_{\theta}
\label{eq_Cr_projection}
\end{equation}
averaged over all lattice sites (angles $\theta$) at the same (or similar) $r$ with $p=0,1,2,\ldots$ 
Due to the $x\leftrightarrow y$-symmetry only even $p=0,2,4,\ldots$ are allowed.
We focus here on $p=0$ (``isotropic projection") 
and $p=2$ (``anisotropic projection") \cite{spmP3}.
If not stated otherwise $p=0$ is assumed.

\subsection{$V$-dependence of observables}
\label{pre_Cr2OV}

As explained in the Introduction it is a general problem to explain
or predict the system-size dependence of an observable $\Pcal(V)$
for asymptotically large volumes $V$. The idea is to express 
$\Pcal(V)^2 = \Eop^{\rvec} C[f](\rvec)$
as an average of a suitable correlation function $C[f](\rvec)$ of a field $f$
which can be independently obtained numerically or understood on theoretical grounds. 
Using the isotropic ($p=0$) projection $C[f](r)$ of $C[f](\rvec)$ we have 
\begin{equation}
V \ \Pcal(V)^2 \approx I(V) \equiv \int \ \ddiff r  \ r^{d-1} \ C[f](r)
\label{eq_Pcal2_interms_Cr}
\end{equation}
in $d$ dimensions. Let us write $\Pcal(V) \sim 1/V^{\gamma}$ for large $V$ using
the phenomenological exponent $\gamma$. Several cases are important.
If $C[f](r)$ vanished more rapidly then $1/r^d$, the integral $I(V)$ is dominated 
by its lower bound and $V \Pcal^2(V)$ becomes constant, hence, $\gamma=1/2$
if the lower bound does not explicitly dependent on $V$.
If on the other hand $C[f](r) \simeq c_0 > 0$ for large $r$ with $c_0$ being a constant, 
$I(V) \simeq c_0 V$ for large $V$ and, hence, 
$\gamma=0$ if $c_0$ is $V$-independent. 
More generally, LSM-C (Table~\ref{tab_model}) illustrates power-law correlations with
\begin{equation}
C[f](r) \simeq c_0/r^{\alpha} \mbox{ for } 1 \ll r \ll L/2
\label{eq_LSMC}
\end{equation}
with $c_0$ being a $V$-independent constant.
While (as already said) $\gamma=1/2$ for $\alpha > d$,
this implies $\gamma = \alpha/2d < 1/2$ for long-range correlations ($\alpha < d$), 
i.e. $\gamma \to 0$ for $\alpha \to 0$. 
Finally, we note that the intermediate case with $\alpha=d$ yields the logarithmic relation 
\begin{equation}
\Pcal(V) \simeq \sqrt{(c_1+c_2 \ln(V))/V} \mbox{ for } V \to \infty
\label{eq_log_behav}
\end{equation}
with $c_1$ and $c_2>0$ being again $V$-independent constants.

\subsection{Imposing $C[a](\rvec)$ and $C[b](\rvec)$}
\label{pre_CarCbr}

\begin{figure}[t]
\centerline{\resizebox{.90\columnwidth}{!}{\includegraphics*{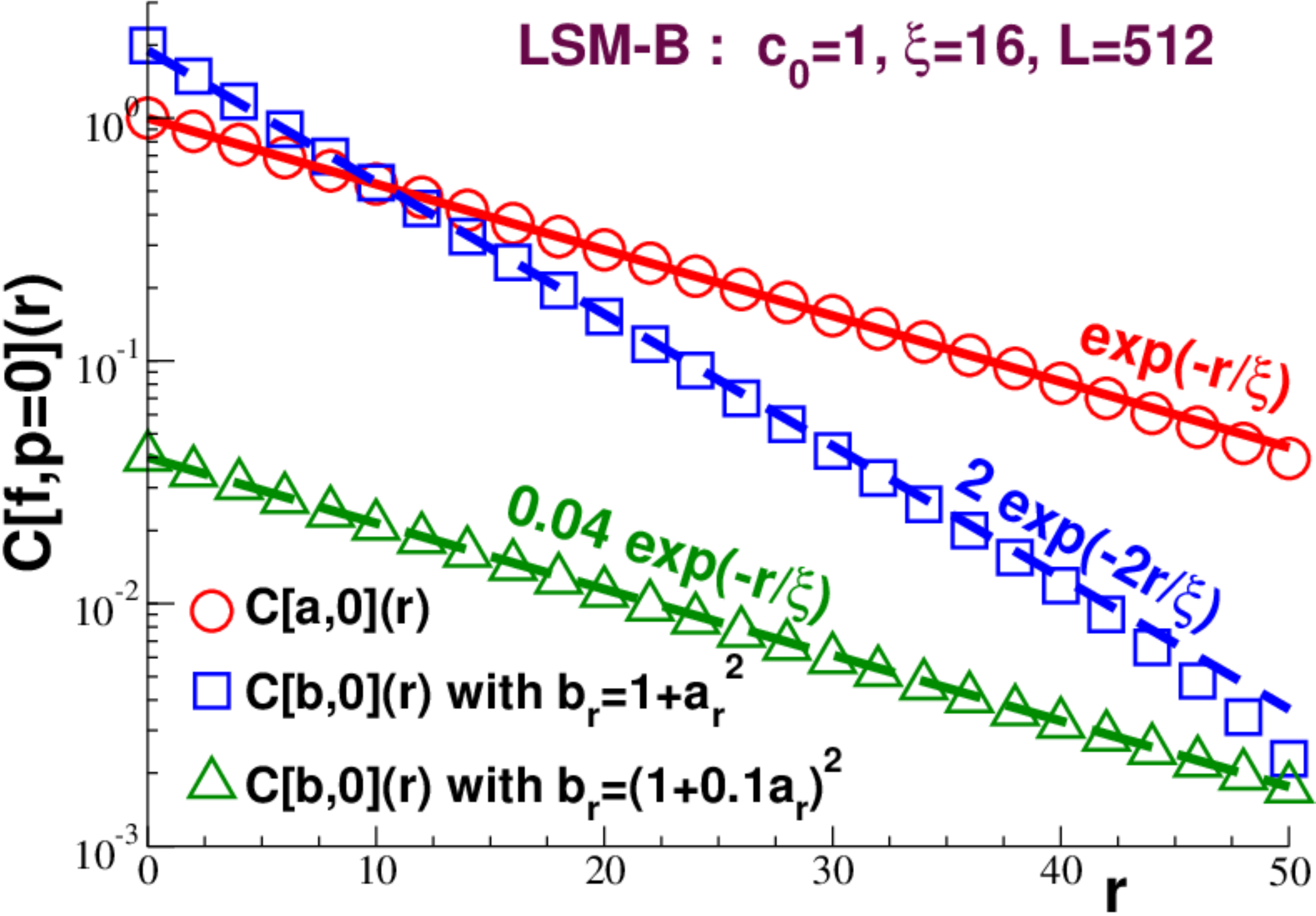}}}
\caption{Isotropically averaged ($p=0$) correlation functions $C[f,p=0](r)$ for 
LSM-B with $c_0=1$, $\xi=16$ and $L=512$ characterizing the quenched fields 
$f=a$ and $f=b$ for all $J$. 
Note that $C[a,0](r)=\exp(-r/\xi)$ as imposed (bold solid line).
For $f=b$ we compare the two closures $\br=1+\ar^2$ (squares) and $\br =(1+0.1\ar)^2$.
The dashed lines indicate the expected behavior for $L \gg \xi$.
}
\label{fig_CarCbr_xi16_caseB}
\end{figure}

\begin{figure}[t]
\centerline{\resizebox{.90\columnwidth}{!}{\includegraphics*{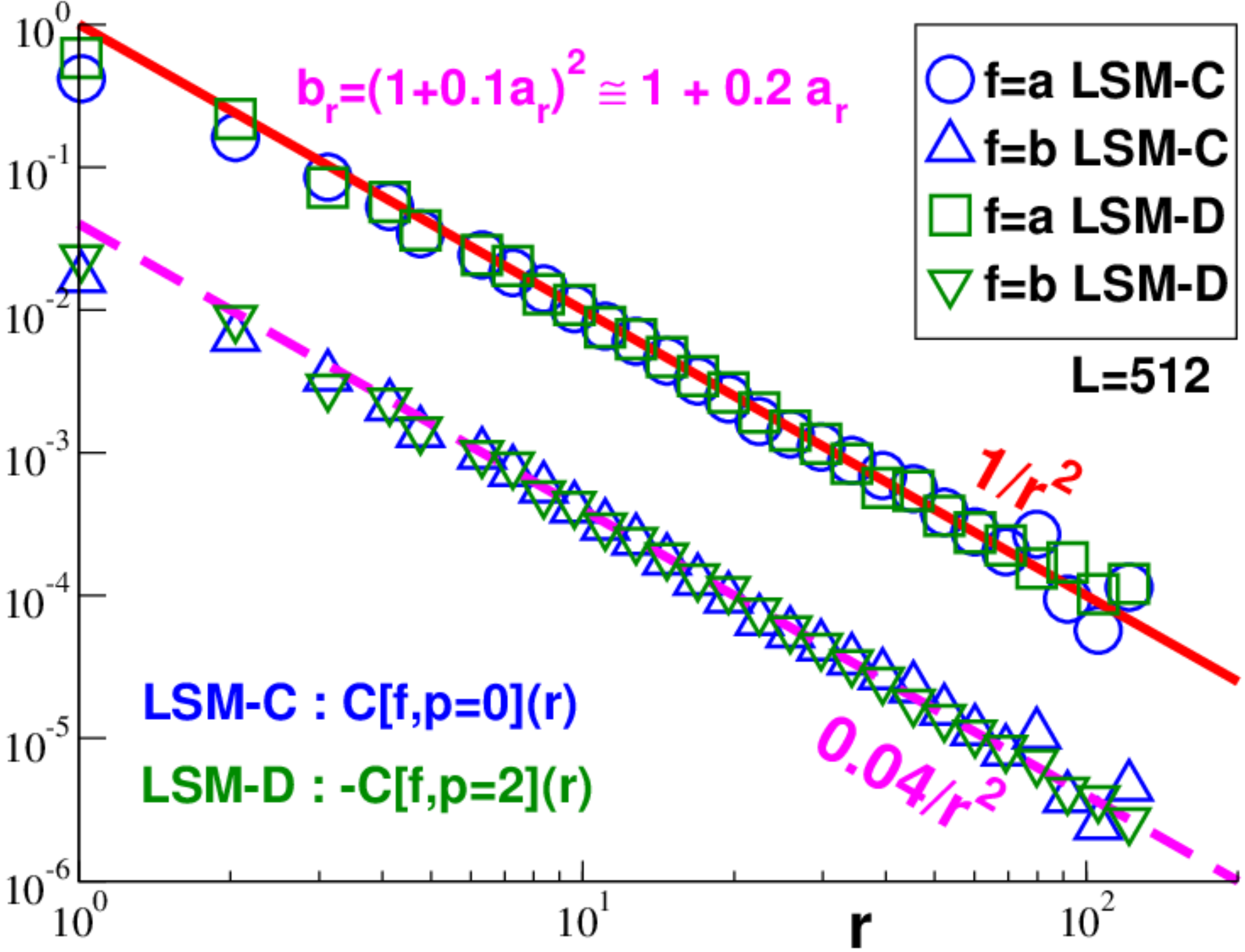}}}
\caption{Double-logarithmic representation of $C[f,p=0](r)$ for LSM-C and 
$-C[f,p=2](r)$ for LSM-D for the two quenched fields $f=a$ and $f=b$
assuming that $\br = (1+0.1 \ar)^2$.
}
\label{fig_CarCbr_log}
\end{figure}

As indicated in Table~\ref{tab_model} the frozen fields $f=a$ and $f=b$ of LSM-A,
our simplest LSM variant, are spatially decorrelated and uniformly
distributed random variables. 
In all other considered cases these fields are spatially correlated
as shown in Fig.~\ref{fig_CarCbr_xi16_caseB} for LSM-B
and in Fig.~\ref{fig_CarCbr_log} for LSM-C and LSM-D.
We explain here how this is done. We remind that, quite generally,
a spatially correlated field with $C[f](\rvec) = \Kimp(\rvec)$ 
is generated by setting
\cite{FractalConcepts,spmP3}
\begin{equation}
\fq = \sqrt{\Kimp(\qvec) \Nr} \ \gq \mbox{ with } \gq = \Fcal\{ \gr \}
\label{eq_fq_correlated}
\end{equation}
being the Fourier transform of a (decorrelated) random Gaussian field $\gr \in \Ncal(0,1)$ 
of zero mean and unit variance.
As a consequence
\begin{equation}
\la |\fq|^2 \ra = \Kimp(\qvec)\Nr \la g_{\qvec} g_{-\qvec} \ra = \Kimp(\qvec),
\label{fq_correlated2}
\end{equation}
i.e. according to the WKT we have $C[f](\rvec)=\Kimp(\rvec)$ 
upon inverse Fourier transform back to real space.\footnote{Using
Parseval's theorem it is seen that $\langle g_{\qvec} g_{-\qvec} \rangle=1/\Nr$.}
It is assumed here that (in addition of being even and periodic functions) 
the imposed $\Kimp(\qvec)$ must be for all $\qvec$ both real and positive 
in agreement with Eq.~(\ref{eq_WKT}). 
If $\Kimp$ is known (stated) in real space, it may be thus necessary to regularize
the desired relation. For instance, the power law $\Kimp(\rvec) = c_0/r^{\alpha}$
must be changed to 
\begin{equation}
\Kimp(\rvec) = c_0 (1+r^2)^{-\alpha/2} 
\label{eq_Kimpr_alpha}
\end{equation}
to avoid the singularity at the origin \cite{FractalConcepts}.
This is done, e.g., with $c_0=1$ and $\alpha=2$ for $f=a$ of LSM-C 
as shown in Fig.~\ref{fig_CarCbr_log}.
To avoid such numerical problems $\Kimp(\qvec)$ is best directly imposed in reciprocal space.
This is the case for $f=a$ and $f=b$ of LSM-D
where we impose the anisotropic correlation function 
\begin{equation}
\Kimp(\qvec) = 8\pi \frac{c_0}{\Nr} \ (\qx \qy/q^2)^2 
\label{eq_Kimpq_aniso}
\end{equation}
($\qx$ and $\qy$ being the discrete components of the wavevector)
motivated by the theoretically predicted shear-stress correlations 
of viscoelastic liquids and glasses \cite{Fuchs17,Fuchs18,Fuchs19,lyuda18}.
As may be seen for $f=a$ of LSM-D in Fig.~\ref{fig_CarCbr_log} (squares)
Eq.~(\ref{eq_Kimpq_aniso}) thus leads in real space to \cite{lyuda18}
\begin{equation}
C(\rvec) = \Kimp(\rvec) = -2c_0 \cos(4\theta)/r^2 \mbox{ for } r > 0,
\label{eq_Cfr_modelD}
\end{equation}
i.e. the isotropic average vanishes (not shown)
and the anisotropic average $C[f,p=2](r) = -c_0/r^2$ is long-ranged (bold solid line).

An additional problem arises for the correlated $f=b$
since all inverse spring constants $\br$ must be positive. To impose this
we generate first an auxiliary field $\ccr$ using the above scheme.
In all cases presented here $\ccr=\ar$.
From this $\br$ is obtained by setting, e.g., $\br = 1 + \ccr^2$.
As seen in Fig.~\ref{fig_CarCbr_xi16_caseB} this ``closure" leads for LSM-B to 
$C[b](\rvec) \approx 2 \exp(-2r/\xi)$ for $L \gg \xi$ 
as can be also proved theoretically. Another possibility is to set 
\begin{equation}
\br = (1+ \lambda \ccr)^2 \approx 1 + 2\lambda \ccr \mbox{ for } |\lambda| \ll 1.
\label{eq_br_2nd_closure}
\end{equation}
With $C[c](\rvec)$ being the correlation function of the auxiliary variable this implies 
to leading order 
\begin{equation}
C[b](\rvec) \approx (2\lambda)^2 C[c](\rvec) \mbox{ for } |\lambda| \ll 1
\label{eq_Cb_2nd_closure}
\end{equation}
which is merely a shift in logarithmic coordinates.
That this works well can be seen (triangles) in Fig.~\ref{fig_CarCbr_xi16_caseB}
for LSM-B and in Fig.~\ref{fig_CarCbr_log} for LSM-C and LSM-D.

\section{Stochastic processes, observables and corresponding fields}
\label{spf}

\subsection{Time-series and associated local fields}
\label{spf_x}

It is common to characterize a stochastic process $x(\tau)$
using ensembles $\{\xbf\}$ of discrete time-series
\begin{equation}
\xbf = \{x_t=x(\tau = t \tincr),t=1,\ldots,\Nt\}
\label{eq_xbf_def}
\end{equation}
with $t$ being the discrete time,
$\tincr$ the time interval between the equidistant measurements and 
$\tsamp = \Nt \tincr$ the available ``sampling time".
We assume that the global stochastic process is a $d$-dimensional volume average
\begin{equation}
x_t = \Eop^{\rvec} x_{t\rvec} \approx \frac{1}{V} \int \ddiff \rvec \ x_{t\rvec}
\label{eq_xtr}
\end{equation}
over a discrete field $x_{t\rvec}$ of same dimension.
As a specific example we consider the spatial average $x_t=\Eop^{\rvec} x_{t\rvec}$
of the LSM spring lengths $x_{t\rvec}$ (cf. Sec.~\ref{pre_LSM}).
It is useful to directly measure $x_{t\qvec} = \Fcal\{x_{t\rvec}\}$ in reciprocal space. 
Since we consider stochastic processes in non-ergodic systems $\xbf_{ck}$, $x_{ckt}$,
$x_{ckt\rvec}$ and $x_{ckt\qvec}$ are additionally characterized by the index $c$ of the independent
configuration and the index $k$ of the time-series of a given $c$. 

\subsection{$t$-averaged observables and fields}
\label{spf_Ox}

Importantly, it is generally not possible to store all sets of
time-series $\xbf$ and associated fields but one normally only computes
and stores functionals (moments) $\Ocal[\xbf]$ of each
time-series, called here ``$t$-averages" or ``observables".
The two observables we shall focus on are
the arithmetic mean
\begin{equation}
\Ocal[\xbf] = \Om[\xbf] \equiv \Eop^t x_t
\label{eq_Om_def}
\end{equation}
and the empirical variance
\begin{equation}
\Ocal[\xbf] = \Ov[\xbf] \equiv \beta V \ \Vop^t x_t
\label{eq_Ov_def}
\end{equation}
with $\beta=1/T$ being the inverse temperature ($\kB=1$).
The prefactor $\beta V$, introduced for consistency with previous work \cite{spmP1,spmP2,spmP3},
is natural for stochastic processes $x_t$ corresponding to intensive thermodynamic variables
\cite{Lebowitz67}.\footnote{In this case $v[\xbf]$ has the dimension of a (free) energy
density just as the stress (pressure) of the system.}
We often write below compactly $\Ocal_{ck}=\Ocal[\xbf_{ck}]$. 

As already pointed out in the Introduction, 
we assume that, as the stochastic process $x_t$,
also the observables $\Ocal$ may be written as linear volume averages 
$\Ocal[\xbf] = \Eop^{\rvec} \Ocal_{\rvec}$ of local contributions $\Ocal_{\rvec}$.
For $\Ocal[\xbf]=m[\xbf]$ these local contributions are given by 
$\Ocal_{\rvec} = m_{\rvec} \equiv \Eop^t x_{t\rvec}$.
Importantly, it is also possible to write the $t$-averaged variance as
$v[\xbf] = \Eop^{\rvec} \vr$ defining the ``local variance"
\begin{eqnarray}
\vr & \equiv & \beta V \  \Eop^t (x_{t\rvec}-x_{\rvec}) (x_t-x) \nonumber \\
& = & \beta V \ \left(\Eop^t x_{t\rvec}x_t - x_{\rvec} x  \right)
\label{eq_Ovr_def}
\end{eqnarray}
with $x_{\rvec} = \Eop^t x_{t\rvec}$ and $x = \Eop^{\rvec} x_{\rvec}$.
Strictly speaking, $\vr$ is a ``co-variance" correlating the local field to the total average. 
Such local variances appear in the stress-fluctuation formulae for 
local elastic moduli \cite{Lutsko88,Lutsko89,Barrat06}.\footnote{The covariance $\vr$ 
must be distinguished from the purely local variance
$\tilde{v}_{\rvec} = \beta V \Eop^t (x_{t\rvec}-x_{\rvec})^2$.}
For numerical reasons it is convenient to compute the local fields 
$\mr$ and $\vr$ in Fourier space from $x_{t\qvec}$ using $\mq= \Eop^t x_{t\qvec}$ and 
$\vq = \beta V \Eop^t (x_{t\qvec}-x_{\qvec}) (x_t-x)$
with $x=x_{\qvec=\bfzero}$.

We remark finally that for the LSM versions with no or weak interactions between neighboring sites 
we have quite generally\footnote{To show the second relation it is used that the $\xr-\ar$ are
decorrelated for $J \to 0$ albeit their first and second moments may be correlated.
$\vr \to \br$ holds for all $V$ and $\beta$ due to prefactor $\beta V$ in the definition of $\vr$.}
\begin{equation}
\mr \to \ar, \vr \to \br \mbox{ for } J \to 0
\mbox{ and } \Nt \to \infty.
\label{eq_mrar_vrbr}
\end{equation}
In other words, since we know $C[a](\rvec)$ and $C[b](\rvec)$ by construction,
Eq.~(\ref{eq_mrar_vrbr}) determines (for the specified limits)
the spatial correlations for the local fields $\mr$ and $\vr$.

\section{Global properties}
\label{glob}

\subsection{Reminder of recent work}
\label{glob_reminder}

Summarizing recent work \cite{spmP2,spmP3} we discuss now several general properties 
of expectation values and variances of observables $\Ocal_{ck} \equiv \Ocal[\xbf_{ck}]$
in non-ergodic systems.
We focus first on the dependences on the number $\Nc$ of independent configurations $c$
and the number $\Nk$ of time-series $k$ for each $c$ and 
discuss then the dependences on sampling time $\tsamp$ and volume $V$.

The first point to be made is that the total average $\Ocal(\Nc,\Nk)$ of the
$\Ocal_{ck}$ can be obtained equivalently by
\begin{equation}
\Ocal(\Nc,\Nk) = \Eop^c \Eop^k \Ocal_{ck} = \Eop^k \Eop^c \Ocal_{ck} = \Eop^l \Ocal_l,
\label{eq_EcEk_commute}
\end{equation}i.e. $c$- and $k$-averages commute and for such ``simple averages" \cite{spmP2}
the two indices $c$ and $k$ can be ``lumped" together in one index $l$
with $\Nl = \Nc \Nk$ as indicated by the last sum.
The order of averaging matters, however, for the variance of $\Ocal_{ck}$
for which three {\em different} definitions are relevant:
\begin{eqnarray}
\dOtot(\Nc,\Nk) & \equiv & \Vop^l \Ocal_l, \label{eq_dOtot_def} \\
\dOint(\Nc,\Nk) & \equiv & \Eop^c \Vop^k \Ocal_{ck}  \mbox{ and }
\label{eq_dOint_def} \\
\dOext(\Nc,\Nk) & \equiv & \Vop^c \Eop^k \Ocal_{ck}.
\label{eq_dOext_def}
\end{eqnarray}
As shown in Ref.~\cite{spmP2}, with these definitions Eq.~(\ref{eq_intro_dOtot}) 
exactly holds.
The ``total variance" $\dOtot(\Nc,\Nk)$ is the standard commonly computed variance \cite{lyuda19a,spmP1,Procaccia16}.
We emphasize that $\dOtot(\Nc,\Nk)$ is again a ``simple average", 
i.e. all time-series $\xbf_{ck}$ are lumped together (index $l$) as for the average $\Ocal(\Nc,\Nk)$,
Eq.~(\ref{eq_EcEk_commute}), while the order of the $c$- and $k$-averaging matters for the 
``internal variance" $\dOint(\Nc,\Nk)$ and the ``external variance" $\dOext(\Nc,\Nk)$.

Let us assume next that $\Nc$ becomes arbitrarily large.
Importantly, the large-$\Nc$ limits $\Ocal$ and $\sOtot$ of $\Ocal(\Nc,\Nk)$ and $\sOtot(\Nc,\Nk)$ 
do neither depend on $\Nc$ nor on $\Nk$ and may, especially, also be computed by using only 
{\em one} time-series for each configuration ($\Nk=1$).
At variance with this, internal and external variances still depend on $\Nk$,
i.e. $\sOint(\Nc,\Nk) \to \sOint(\Nk)$ and $\sOext(\Nc,\Nk) \to \sOext(\Nk)$ 
in general for $\Nc \to \infty$. 
Note that 
\begin{equation}
\sOint(\Nk) \to 0, \sOext(\Nk) \to \sOtot \mbox{ for } \Nk \to 1.
\label{eq_glob_Nkone}
\end{equation}
For large spacer times $\tspacer \gg \taubasin$ between time-series
the $\Nk$-dependence is given using Eq.~(\ref{eq_Vop_Nl}) by \cite{spmP2}
\begin{eqnarray}
\dOint(\Nk) & \simeq & \left(1 - \frac{1}{\Nk} \right) \ \dOint
\label{eq_dOint_Nk} \\
\dOext(\Nk) & \simeq & \dOext + \frac{1}{\Nk} \ \dOint
\label{eq_dOext_Nk}
\end{eqnarray}
where $\sOint$ and $\sOext$ without the argument $\Nk$ stand for the limit $\Nk\to\infty$. 
Using these relations it is possible (with a bit of care as discussed in Sec.~\ref{glob_LSM}) 
to extrapolate internal and external variances measured at finite $\Nk$ to the respective large-$\Nk$ limits.
We focus below on properties corresponding to the large-$\Nc$ and large-$\Nk$ limits.

The above properties may depend additionally on the sampling time $\tsamp$ and the volume $V$. 
The first dependence is relevant for all considered properties below and 
around the basin relaxation time $\taubasin$.
In this work we shall mainly focus on the opposite large-$\tsamp$ limit ($\tsamp \gg \taubasin$).
In this limit the typical $k$-averaged $\Ocal_{ck}$ become $\tsamp$-independent. Hence,
$\Ocal(\tsamp,V) \to \Ocal(V)$ and $\sOext(\tsamp,V) \to \Snonerg(V)$ with $\Snonerg(V)$ 
being the ``non-ergodicity parameter" defined in the Introduction, 
Eq.~(\ref{eq_intro_Snonerg}).\footnote{Following Ref.~\cite{spmP2}
{\em one} simple possibility to characterize $\taubasin$ is to set $\Ocal(\tsamp=\taubasin,V) = f \Ocal(V)$ 
using a fixed fraction $f$ close to unity. We use $f=0.95$.}
At variance with this $\sOint(\tsamp,V)$ remains $\tsamp$-dependent decaying as
\begin{equation}
\sOint(\tsamp,V) \propto \sqrt{\taubasin/\tsamp} \mbox{ for } \tsamp \gg \taubasin
\label{eq_sOint_large_tsamp}
\end{equation}
since we average over $\tsamp/\taubasin$ independent subintervals \cite{spmP2}.
Let us define the ``non-ergodicity time" $\Tnonerg(V) \gg \taubasin$ by $\sOint(\Tnonerg,V) = \Snonerg(V)$. 
$\sOtot(\tsamp,V)$ is dominated by the internal fluctuations, Eq.~(\ref{eq_sOint_large_tsamp}),
for $\tsamp \ll \Tnonerg(V)$ while 
\begin{equation}
\sOtot(\tsamp,V) \to \sOext(\tsamp,V) \approx \Snonerg(V) 
\label{eq_sOtot_large_tsamp}
\end{equation}
in the large-$\tsamp$ limit ($\tsamp \gg \Tnonerg(V)$).
If only the standard total variance is probed the non-ergodicity of the system 
may remain unnoticed for $\tsamp \ll \Tnonerg(V)$.
As further emphasized below, it is then necessary to systematically
check the $\tsamp$-dependence of $\sOtot(\tsamp,V)$ and to carefully extrapolate
to $\tsamp \to \infty$ \cite{spmP2,spmP3}.
The volume dependence will be addressed in more detail in Sec.~\ref{glob_Vol}.
As a consequence $\Tnonerg(V)$ is found to strong\-ly {\em increase} with $V$ 
since $\Snonerg(V)/\sOint(\tsamp,V)$ quite generally decreases with the system size.
Assuming the latter ratio to decay as $1/V^{\gamma}$ this implies $\Tnonerg(V) \propto V^{2\gamma}$.
The determination of $\Snonerg(V)$ by means Eq.~(\ref{eq_sOtot_large_tsamp}) thus becomes 
increasingly difficult.

\subsection{Focus and examples}
\label{glob_LSM}

We focus now on $\Ocal[\xbf]=v[\xbf]$ and the corresponding
expectation value $v(\tsamp,V)$, Eq.~(\ref{eq_EcEk_commute}), and the three 
associated standard deviations $\svtot(\tsamp,V)$, $\svint(\tsamp,V)$ and $\svext(\tsamp,V)$
determined according to Eqs.~(\ref{eq_dOtot_def}-\ref{eq_dOext_def}).
We begin by discussing $\tsamp$-effects (Sec.~\ref{glob_tsamp}) and 
turn then to the $V$-dependence of these properties (Sec.~\ref{glob_Vol}).
We illustrate various points made above by means of MC simulations
of the LSM introduced in Sec.~\ref{pre}.
For all cases we have $T=1$, $\tincr=10$, $\Nc=200$ and at least $\Nk=100$. 
Using Eq.~(\ref{eq_dOint_Nk}) and Eq.~(\ref{eq_dOext_Nk}) 
we extrapolate to $\Nk \to \infty$.
$\svint(\tsamp,V,\Nk)$ can readily be extrapolated
to $\svint(\tsamp,V)$ even using small $\Nk < 10$ as discussed in Ref.~\cite{spmP2}.
At variance with this the extrapolation from $\svext(\tsamp,V,\Nk)$ to $\svext(\tsamp,V)$ 
turns out to be inaccurate if the correction term 
\begin{equation}
\frac{1}{\Nk} \dvint(\tsamp,V) \propto \frac{\taubasin}{\Nk\tsamp}
\mbox{ for } \tsamp \gg \taubasin
\label{eq_dvintNk_correction}
\end{equation}
in Eq.~(\ref{eq_dOext_Nk}) is not small compared to $\dvext(\tsamp,V,\Nk)$.
This matters especially for $\tsamp \ll 10^3$ and $J > 0.1$ 
when the stochastic process becomes slow, increasing thus $\taubasin(J)$.
Occasionally, we have thus been forced to use $\Nk=1000$.\footnote{A 
spacer time interval $\tspacer \approx \tsamp$ is used between each measured 
time series of length $\tsamp$. It may have been more efficient to use instead 
$\tspacer \approx \max(\tsamp,\taubasin(J))$ to make the only asymptotically exact 
Eq.~(\ref{eq_dOext_Nk}) applicable for smaller $\Nk$.}

\subsection{Sampling time dependence}
\label{glob_tsamp}

\begin{figure}[t]
\centerline{\resizebox{.90\columnwidth}{!}{\includegraphics*{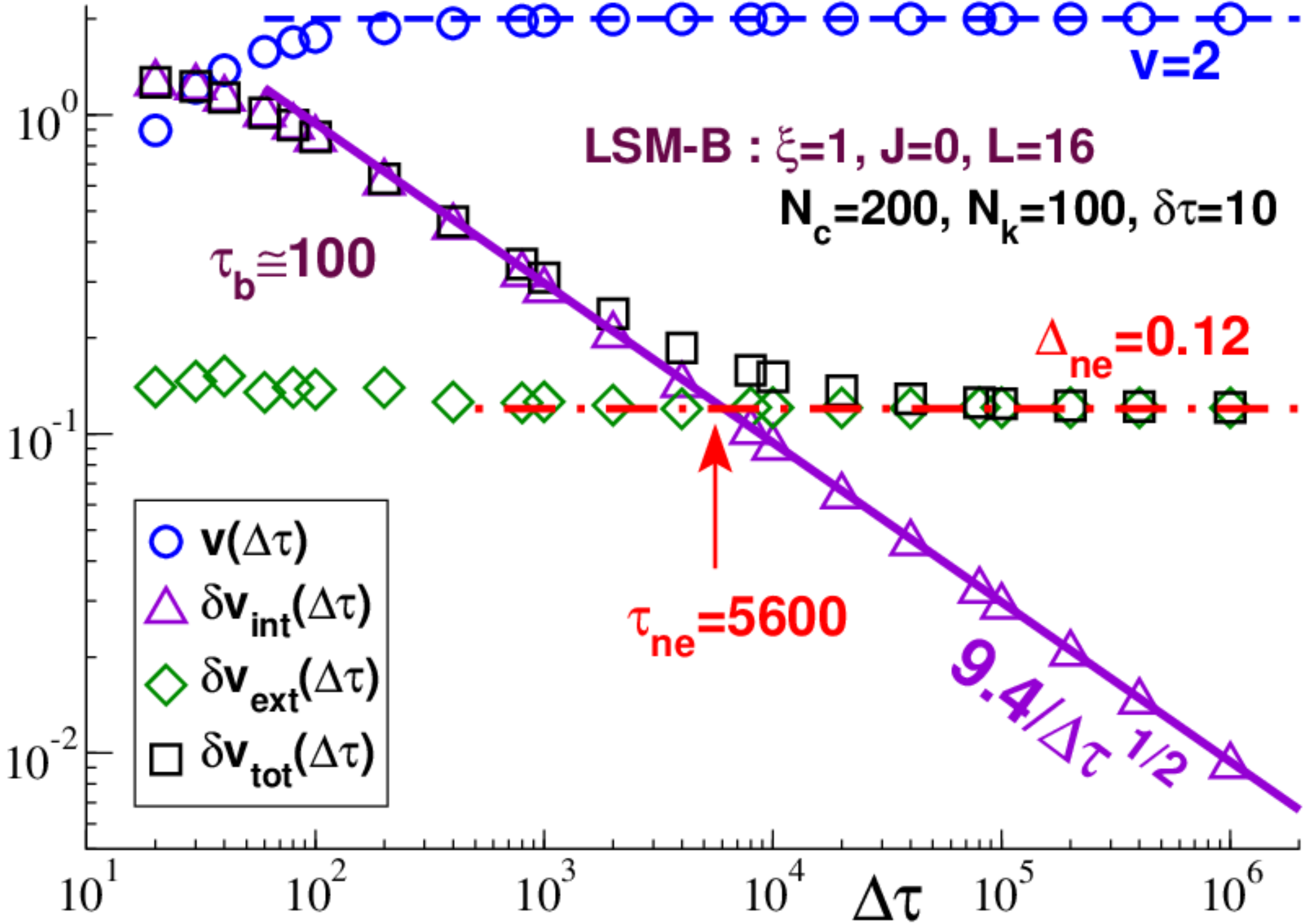}}}
\caption{Sampling time dependence of $v(\tsamp)$, $\svint(\tsamp)$, $\svext(\tsamp)$
and $\svtot(\tsamp)$ for LSM-B with $\xi=1$, $J=0$ and $L=16$.
It is seen that $\svint(\tsamp) \propto 1/\sqrt{\tsamp}$ (bold solid line)
while all other properties become constant for large $\tsamp$.
Note that $\svext(\tsamp) \approx \Snonerg$ much faster than $\svtot(\tsamp)$.
}
\label{fig_vdv_Dt_J0_caseB}
\end{figure}

As a generic example we present in Fig.~\ref{fig_vdv_Dt_J0_caseB} 
data obtained for LSM-B (exponentially decaying $a$- and $b$-fields)
using a correlation length $\xi=1$ and a small simulation box with $L=16$.\footnote{We
often suppress in this subsection the possible additional $V$-dependences, i.e.
we write, e.g., $v(\tsamp)$ instead of $v(\tsamp,V)$.}
All interactions between neighboring springs $\xr$ are switched off ($J=0$).
As can be seen, $v(\tsamp)$, $\svint(\tsamp)$ and $\svext(\tsamp)$
reach rapidly for $\tsamp \gg \taubasin \approx 100$ the asymptotic
behavior expected from the general discussion in Sec.~\ref{glob_reminder},
i.e. $v(\tsamp)$ and $\svext(\tsamp)$ approach the respective constants
$v$ and $\svext = \Snonerg$ while $\svint(\tsamp) \propto 1/\sqrt{\tsamp}$ (bold solid line).
Importantly, due to Eq.~(\ref{eq_intro_dOtot}) 
\begin{equation}
\svtot(\tsamp,V) = \sqrt{\dvint(\tsamp,V)+\dvext(\tsamp,V)}
\label{eq_svtot}
\end{equation}
approaches $\Snonerg$ much later than $\svext(\tsamp)$,
i.e. only for $\tsamp \gg \Tnonerg \approx 5600 \gg \taubasin$.
For this reason it is problematic to determine $\Snonerg$ only by 
measuring $\svtot(\tsamp)$ for {\em one} sampling time.\footnote{If $\svtot(\tsamp)$
is known for a broad range of $\tsamp$ one may plot $\svtot(\tsamp)$
as a function of $1/\sqrt{\tsamp}$ in linear coordinates. 
$\Snonerg$ may then be obtained for $\Nk=1$ from the intercept of the vertical axis of a 
linear data fit. This procedure allows to avoid the determination of $\svext(\tsamp,\Nk)$.}

\begin{figure}[t]
\centerline{\resizebox{.90\columnwidth}{!}{\includegraphics*{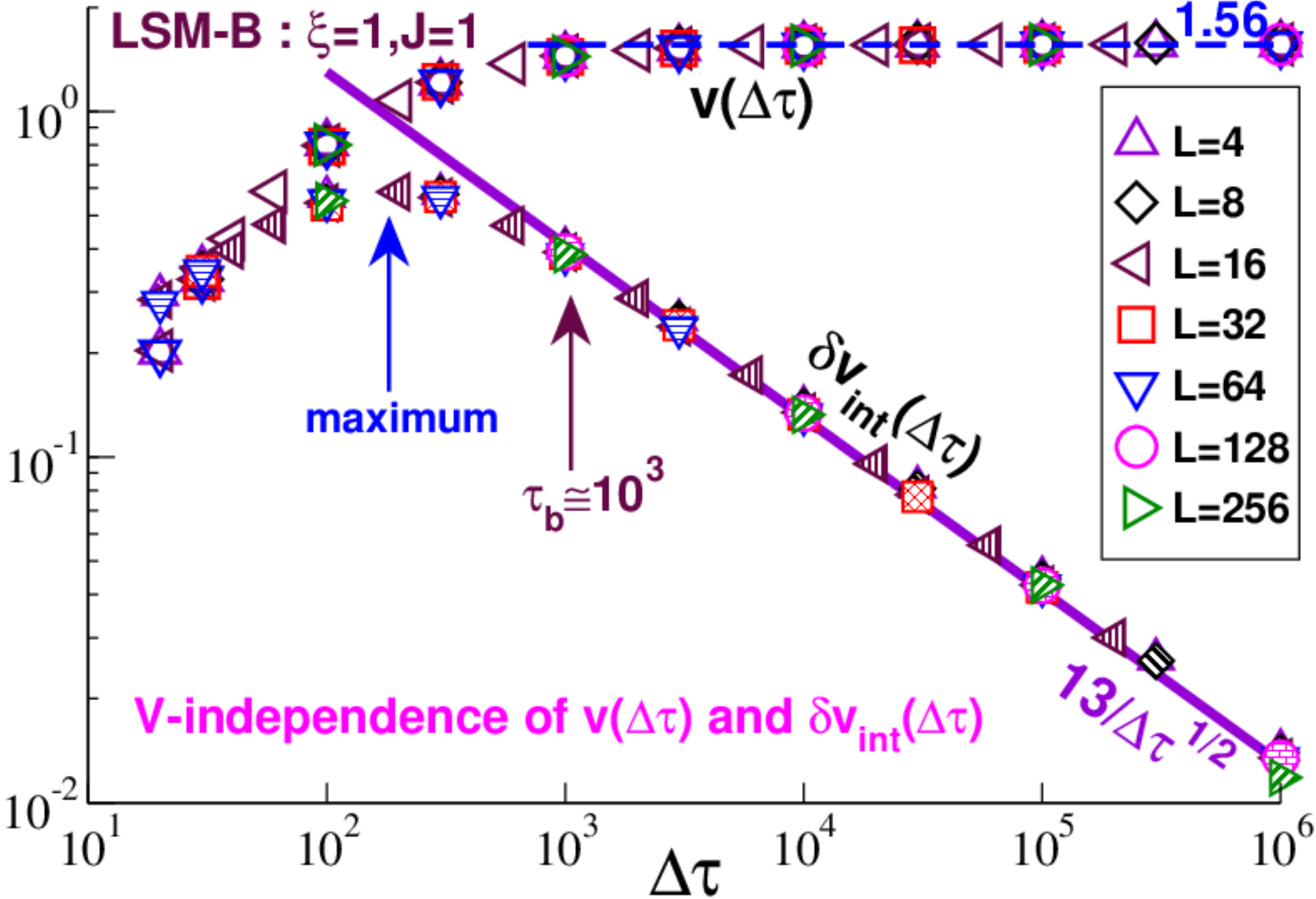}}}
\caption{$v(\tsamp)$ (upper data with open symbols) and $\svint(\tsamp)$ (symbols filled with pattern) 
for LSM-B with $\xi=1$ and $J=1$ for a broad range of $L$ demonstrating the $V$-independence of both properties. 
Increasing $J$ increases $\taubasin$, decreases $v$ (horizontal dashed line)
and increases the power-law amplitude for $\svint(\tsamp) \simeq 1/\sqrt{\tsamp}$ for $\tsamp \gg \taubasin$ (bold solid line).
}
\label{fig_vdv_Dt_J1_caseB}
\end{figure}

Deviations from this generic asymptotic behavior are visible for small $\tsamp$ around and 
below the basin relaxation time $\taubasin$. 
This can be seen from Fig.~\ref{fig_vdv_Dt_J0_caseB} but more clearly from Fig.~\ref{fig_vdv_Dt_J1_caseB} 
where we present $v(\tsamp)$ and $\svint(\tsamp)$ for LSM-B with $J=1$ and 
for a broad range of system-sizes.
Remarkably, $\svint(\tsamp)$ reveals non-monotonic behavior with a maximum below 
the basin relaxation time $\taubasin(J=1) \approx 10^3$.
Being generally due to relaxation processes within each meta-basin this small-$\tsamp$ regime is 
more relevant for more realistic models as discussed elsewhere \cite{lyuda19a,spmP1,spmP2}.
For the present work it is only important to stress that the general
$\tsamp$-dependence of $v(\tsamp)$ and $\svint(\tsamp)$ can be traced back  
to the ``mean-square displacement" (MSD) $h(\tau)$ of the stochastic process.
This is defined by
\begin{equation}
h(\tau= t \tincr) \equiv h_{t=|i-j|} \equiv \frac{\beta V}{2} \la (x_i-x_j)^2 \ra
\label{eq_ht_def}
\end{equation}
averaged over all time entries $i$ and $j$ of a long trajectory with $t=|i-j|$.
For stationary processes \cite{spmP1}
\begin{equation}
v(\tsamp) = \frac{2}{\Nt^2} \sum_{t=1}^{\Nt-1} (\Nt-t)  \ h_t 
\label{eq_vtsamp}
\end{equation}
must hold.\footnote{In
in statistical mechanics Eq.~(\ref{eq_vtsamp}) is closely related to the 
equivalence of the Green-Kubo and the Einstein relations for transport coefficients 
\cite{HansenBook,spmP1,AllenTildesleyBook}.}
The sampling time dependence of $\svint(\tsamp)$ can be understood and described
assuming a stationary {\em Gaussian} stochastic process \cite{lyuda19a,spmP1,spmP2}.
This implies that
\begin{eqnarray}
\dvint(\tsamp)  & = & \dvgauss[h]
\equiv \frac{1}{2\Nt^4} \sum_{i,j,k,l=1}^{\Nt} \ g_{ijkl}^2  
\mbox{ with } 
\label{eq_svint_tsamp} \\
g_{ijkl} & \equiv & (h_{i-j} + h_{k-l}) - (h_{i-l} + h_{j-k}).
\nonumber
\end{eqnarray}
Numerical more convenient alternative representations are given elsewhere \cite{lyuda19a,spmP1}.
By analyzing the functional $\svgauss[h]$ it is seen \cite{lyuda19a,spmP1} 
that while $\svint(\tsamp) \propto 1/\sqrt{\tsamp}$ for (to leading order)
$h(t) \approx const$ for $t \approx \tsamp$, it may become large with
$\svint(\tsamp) \approx v(\tsamp)$ for sampling times $\tsamp$ corresponding 
to a strong change of $h(t\approx \tsamp)$.\footnote{No general relation such as Eq.~(\ref{eq_svint_tsamp})
for $\svint(\tsamp)$ is known at present for $\svext(\tsamp)$.}
We emphasize finally that since $h(\tau)$, $v(\tsamp)$ and $\svint(\tsamp)$
are connected through Eq.~(\ref{eq_vtsamp}) and Eq.~(\ref{eq_svint_tsamp})
all three quantities must have the {\em same} system-size dependence
and this for all times.
That $v(\tsamp)$ and $\svint(\tsamp)$ in Fig.~\ref{fig_vdv_Dt_J1_caseB}
are {\em both} $V$-independent is one consequence.

\subsection{Volume dependence}
\label{glob_Vol}

\begin{figure}[t]
\centerline{\resizebox{.90\columnwidth}{!}{\includegraphics*{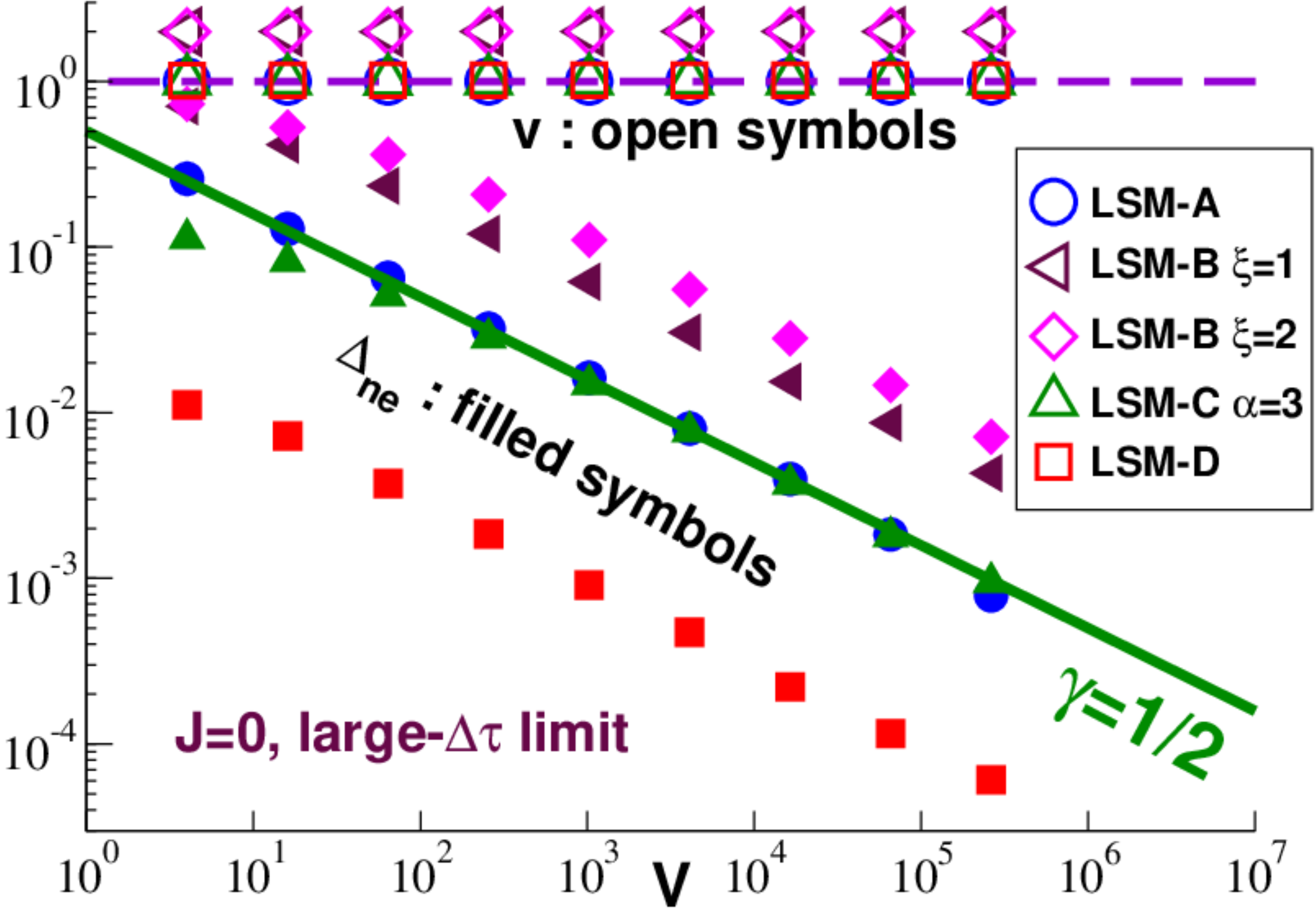}}}
\caption{$V$-dependence for $v$ and $\Snonerg$ (filled symbols)
for LSM-A, LSM-B with $\xi=1$ and $\xi=2$, LSM-C with $\alpha=3$ and LSM-D.
$v$ is always $V$-independent (horizontal dashed line)
and $\Snonerg \propto 1/V^{\gamma}$ with $\gamma=1/2$ (bold solid line)
for the given examples with short-range correlations.
$\Snonerg(V)$ for LSM-D (filled squares) is finite, but much smaller than all other cases.
}
\label{fig_vdvext_V}
\end{figure}

\begin{figure}[t]
\centerline{\resizebox{.90\columnwidth}{!}{\includegraphics*{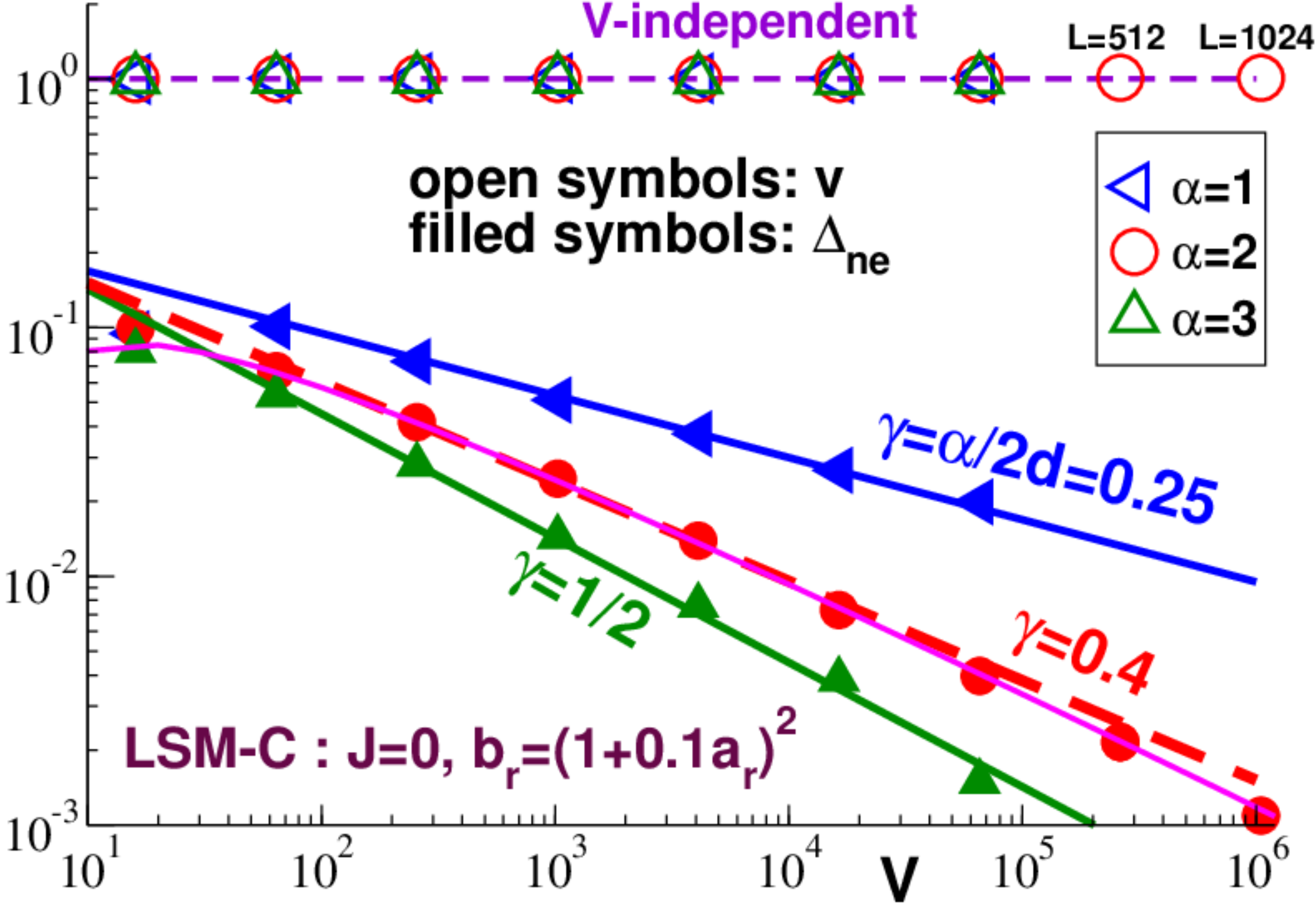}}}
\caption{Volume dependence of $v$ and $\Snonerg$ for LSM-C and three different values of $\alpha$.
We observe short-range behavior with $\gamma=1/2$ for $\alpha > d$,
long range behavior with $\gamma=\alpha/2d$ for $\alpha < d$
and, as expected from Eq.~(\ref{eq_log_behav}), 
logarithmic decay (thin solid line) for the intermediate case with $\alpha=d =2$.
The latter case is well {\em fitted} by an exponent $\gamma=0.4$ (bold dashed line). 
}
\label{fig_dvext_V_caseC}
\end{figure}

\begin{figure}[t]
\centerline{\resizebox{.90\columnwidth}{!}{\includegraphics*{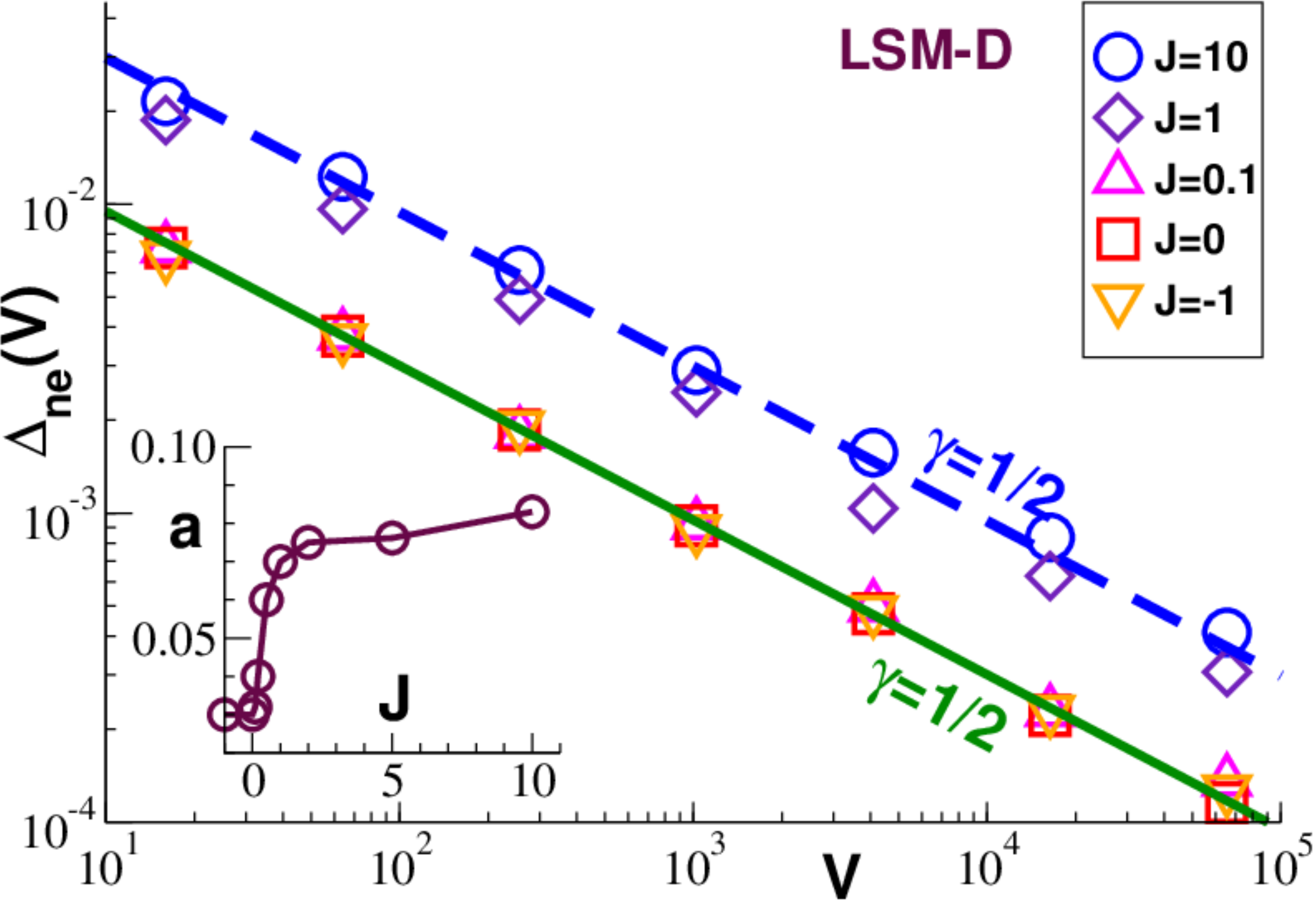}}}
\caption{$\Snonerg(V,J)$ for LSM-D for different $J$.
Main panel: $\gamma=1/2$ holds for all $J$ (lines).
Inset: Power-law amplitude $a(J) \equiv \Snonerg(V,J) V^{1/2}$ {\em vs.} coupling constant $J$.
Apparently, the spring coupling introduces isotropic ($p=0$) correlations
of the $x_{t\rvec}$- and, hence, $v_{\rvec}$-fields which remain, however, short-ranged.
}
\label{fig_dvext_V_caseD}
\end{figure}

We turn now to system-size effects.
Let us focus first on the limit $\tsamp \gg \Tnonerg(V)$
where $\svint(\tsamp)$ becomes negligible and
$\svtot(\tsamp) \approx \svext(\tsamp) \approx \Snonerg$. 
Examples for $v(V)$ and $\Snonerg(V)$ are given for $J=0$
in Fig.~\ref{fig_vdvext_V} and Fig.~\ref{fig_dvext_V_caseC}
and for $\Snonerg(V)$ comparing different $J$ for LSM-D in Fig.~\ref{fig_dvext_V_caseD}.

The first point to be made is that the variance $v$ is always $V$-independent
(as already seen in Fig.~\ref{fig_vdv_Dt_J1_caseB}) 
due to the prefactor $\beta V$ introduced in Eq.~(\ref{eq_Ov_def}).
In fact this scaling is expected to hold for all stochastic processes
describing intensive system properties if the $c$-trajectories are at 
thermal equilibrium in their respective basins \cite{spmP2}.
(Note that each stochastic process is ergodic in its basin.)
Using the standard fluctua\-tion-dissipation relation for the fluctuation 
of intensive thermodynamic variables \cite{Lebowitz67,WXP13,AllenTildesleyBook} 
this implies that $v_c$ does not depend explicitly on $V$
and, hence, neither does $v = \Eop^c v_c$.\footnote{It
is well known that $v_c$ depends on whether the average intensive variable of the
basin is imposed or its conjugated extensive variable \cite{Lebowitz67}.}
This argument even holds for systems with long-range correlations 
if standard thermostatistics can be used for each basin.
This can be seen from the variances $v$ of LSM-C (power-law correlations) 
for exponents $\alpha < d$ as shown in Fig.~\ref{fig_dvext_V_caseC}
for $\alpha=1$. 

Interestingly, the same thermodynamic reasoning {\em cannot} be made for $\Snonerg$.
However, it can be readily demonstrated that quite generally 
$\Snonerg \propto 1/V^{\gamma}$ with $\gamma=1/2$ for systems without
spatial correlations \cite{spmP2}.
This is the case for LSM-A with $J=0$ on which we may focus without loss of generality.
According to Eq.~(\ref{eq_mrar_vrbr}) we have $b_{c\rvec}=v_{c\rvec}$ and
thus $v_c=\Eop^{\rvec} v_{c\rvec}$ is given 
by the spatial average $b_c= \Eop^{\rvec} b_{c\rvec}$.
This implies in turn that $v \equiv \Eop^c v_c = \Eop^c b_c \equiv b$.\footnote{Without 
invoking here thermostatistics this argument demonstrates that $v$ 
must be $V$-independent whenever the spatial correlations are short-ranged.}
To get the variance of the variance $\Vop^c v_c$ one uses again that the variance of the sum of 
stochastic independent variables is the sum of the variances of those variables
\begin{equation}
\Dnonerg = \Vop^c v_c =\Vop^c \left(\frac{1}{\Nr} \sum_{\rvec} b_{c\rvec}\right) = 
\frac{1}{\Nr} \times \underline{\Eop^{\rvec} \Vop^c b_{c\rvec}}
\end{equation}
and the fact that the underlined term does not depend on 
the number of grid sites $\Nr \propto V$ for large systems.
Hence, $\gamma=1/2$.
Naturally, this does not only hold for systems with strictly decorrelated fields but also if 
short-range correlations are present (which may be renormalized away) as confirmed by the 
various additional examples with short-range correlations\footnote{While
``short-range" is often reserved for ultimately exponentially decaying correlation functions,
it is used here also for correlations decaying sufficiently fast 
such that the volume average does not depend on the upper integration boundary $L$.}
presented in Fig.~\ref{fig_vdvext_V}, Fig.~\ref{fig_dvext_V_caseC} 
and Fig.~\ref{fig_dvext_V_caseD}.
(As shown in the latter plot for LSM-D, the coupling parameter $J$ has apparently 
only a weak quantitative effect on the range of the {\em effective} spatial correlations.)
The above argument breaks down, however, if long-range correlations
are present as for the power-law exponent $\alpha=1$ of LSM-C
shown in Fig.~\ref{fig_dvext_V_caseC}.
The observed power law with $\gamma = \alpha/2d$ is, of course,
expected from Sec.~\ref{pre_Cr2OV} as we shall corroborate below in Sec.~\ref{corr}.

\begin{figure}[t]
\centerline{\resizebox{.90\columnwidth}{!}{\includegraphics*{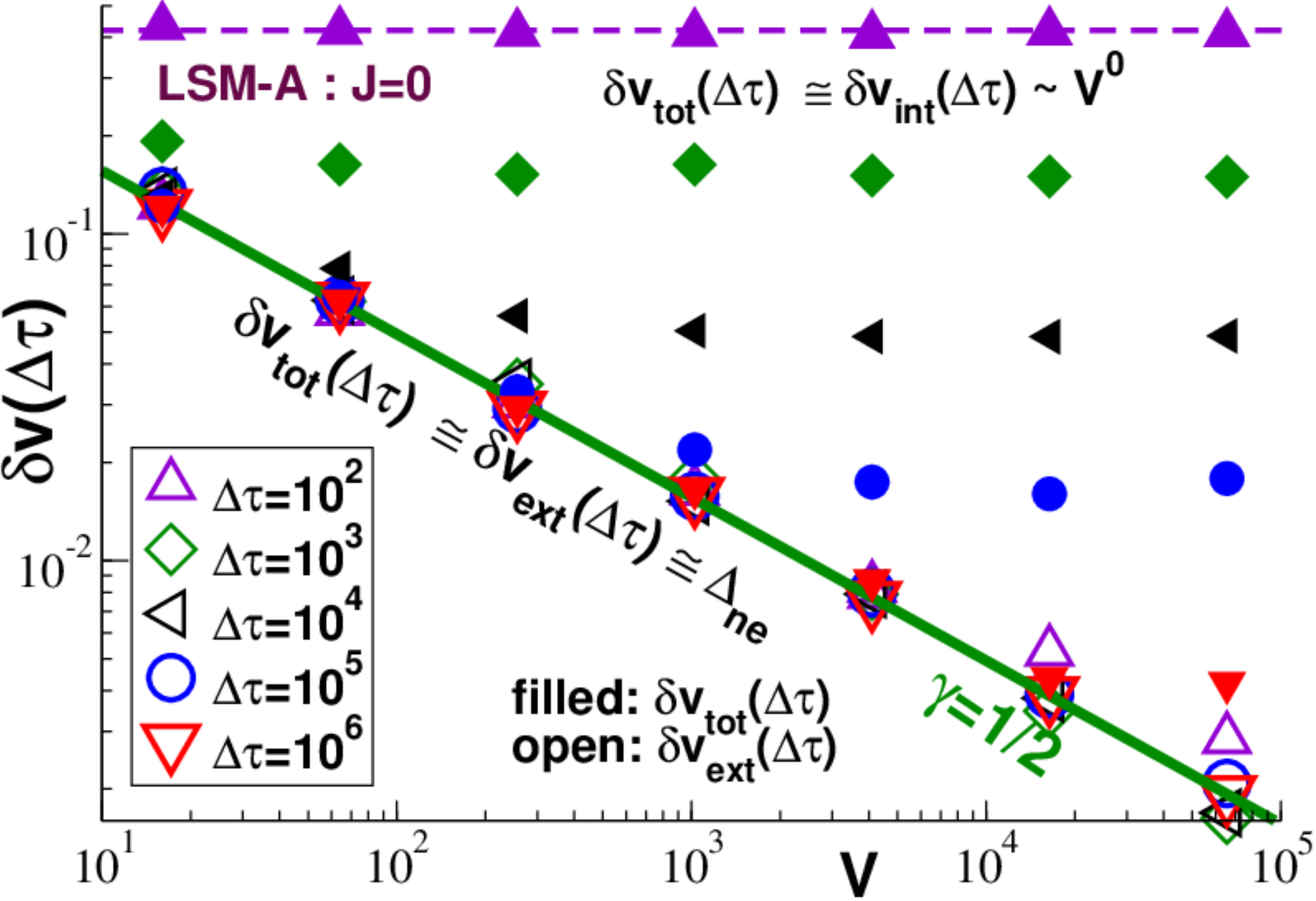}}}
\caption{$V$-dependence of $\svtot(\tsamp)$ and $\svext(\tsamp)$
for LSM-A for different sampling times $\tsamp$ as indicated.
The bold green line indicates the known 
$\Snonerg \propto 1/V^{\gamma}$ with $\gamma=1/2$.
While $\svext(\tsamp,V) \approx \Snonerg(V)$ for all $L$ and $\tsamp \gg \taubasin \approx 10$,
a much slower convergence to this limit is seen for $\svtot(\tsamp,V)$ 
due to the $V$-independent contribution $\svint(\tsamp)$ to $\svtot(\tsamp)$.
}
\label{fig_dvext_V_caseA}
\end{figure}

Let us instead end this paragraph with some comments on the $\tsamp$-dependence 
of the system-size effects. We remind that $v(\tsamp)$, $\svint(\tsamp)$ and $h(t)$ 
are related via Eq.~(\ref{eq_vtsamp}) and Eq.~(\ref{eq_svint_tsamp}). 
In view of the observed $V$-independence of $v$ it is thus not surprising that $v(\tsamp)$ 
and $\svint(\tsamp)$  are found to be $V$-independent for all $\tsamp$
as shown in Fig.~\ref{fig_vdv_Dt_J1_caseB}.
Since $\svint(\tsamp) \propto V^0/\sqrt{\tsamp}$ for $\tsamp \gg \taubasin$
the non-ergodicity crossover time 
$\Tnonerg(V) \simeq V^{2\gamma}$ rapidly increases with $V$. 
This implies that the regime with $\taubasin \ll \tsamp \ll \Tnonerg(V)$ where 
$\svtot(\tsamp) \approx \svint(\tsamp)$ strongly increases with $V$.  
If computed at constant $\tsamp$ as in most computational studies \cite{Procaccia16}, 
$\svtot(\tsamp,V)$ as a function of $V$ must thus become $V$-independent for large $V$. 
This behavior can clearly be seen from the data presented 
in Fig.~\ref{fig_dvext_V_caseA} for LSM-A (filled symbols).\footnote{We
use here $\Nk=1000$ for $\tsamp \le 10^4$ to obtain for $\svext(\tsamp,\Nk)$
a sufficiently accurate $\Nk$-extrapolation $\svext(\tsamp)$ for small sampling times.}
The crossing over from $\svtot(\tsamp,V) \approx \Snonerg(V) \propto 1/V^{\gamma}$ for small $V$
(bold solid line) to $\svtot(\tsamp,V) \approx \svint(\tsamp) \propto V^0$ for large $V$ makes it likely 
that in turn a too small apparent exponent $\gamma$ may be fitted. 
Should $\svext(\tsamp,V)$ not be available one needs at least to compare $\svtot(\tsamp,V)$ 
for several $\tsamp$. Only the $V$-regime where the highest $\tsamp$-data do indeed collapse 
can be used for a fit of the exponent $\gamma$.
Without such a crosscheck all fits claiming an exponent $\gamma < 1/2$
and, hence, (according to the preceding paragraph) long-range spatial correlations
are questionable.

\section{Associated spatial correlation functions}
\label{corr}

\subsection{General relations for non-ergodic systems}
\label{corr_inexto}

As demonstrated in detail in Appendix~\ref{app_corr} the integrals, 
Eqs.~(\ref{eq_intro_Ctotr2dOtot}-\ref{eq_intro_Cextr2dOext}), 
are solved by
\begin{eqnarray}
\Ctot(\qvec) & = & \Eop^l \Khat[\Ocal_{l\qvec}](\qvec) - \Ocal^2\delta_{\qvec\bfzero}
\label{eq_Ctotq_def} \\
\Cint(\qvec) & = & \Eop^c \Eop^k \Khat[\Ockq-\Ocq](\qvec)
\label{eq_Cintq_def} \\
\Cext(\qvec) & = & \Eop^c \Khat[\Ocq](\qvec) - \Ocal^2\delta_{\qvec\bfzero}
\label{eq_Cextq_def} 
\end{eqnarray}
where for numerical convenience we have stated all correlation functions in reciprocal space
(with $\delta_{\qvec\bfzero}$ denoting Kronecker's symbol for the zero-wavevector contribution).
The ``simple average" $\Ctot$ corresponds to the standard commonly measured correlation function. 
The internal correlation function $\Cint$ characterizes the correlations of the difference $\Ockq- \Ocq$
with respect to the $k$-average $\Ocq =\Eop^k \Ockq$.
Moreover, as shown by Eq.~(\ref{eq_Cintr_def_B}) the ``total" correlation function 
$\Ctot$ is the sum of an ``internal" contribution $\Cint$ and an ``external" contribution $\Cext$
\begin{equation}
\Ctot(\qvec) = \Cint(\qvec) + \Cext(\qvec)
\label{eq_intro_Cinexto_q}
\end{equation}
in agreement with Eq.~(\ref{eq_intro_Cinexto}) stated in the Introduction.
 
Just as $\sOtot$, $\sOint$ and $\sOext$ the correlation functions $\Ctot$, $\Cint$ and $\Cext$
depend in general on $\Nc$ and $\Nk$. 
As above in Sec.~\ref{glob_reminder} we assume that $\Nc$ is arbitrarily large.
This implies that
\begin{equation}
\lim_{\Nc \to \infty} \Ctot(\qvec,\Nc,\Nk) = \Ctot(\qvec)
\label{eq_Ctot_Nk}
\end{equation}
and similarly in real space,
i.e. not only the $\Nc$- but also the $\Nk$-dependence drops
out since the total correlation function is a simple $l$-average.
Consistently with Eq.~(\ref{eq_dOint_Nk}) we have for the internal correlation
function
\begin{equation}
\Cint(\qvec,\Nk) \simeq \left(1-\frac{1}{\Nk}\right) \Cint(\qvec)
\label{eq_Cint_Nk}
\end{equation} 
for $\Nc \to \infty$. Using Eq.~(\ref{eq_Ctot_Nk}), Eq.~(\ref{eq_Cint_Nk})
together with Eq.~(\ref{eq_intro_Cinexto}) it is then seen that
\begin{equation}
\Cext(\qvec,\Nk) \simeq \Cext(\qvec) + \frac{1}{\Nk} \Cint(\qvec).
\label{eq_Cext_Nk}
\end{equation}
These two relations should be used to extrapolate for 
the asymptotic $\Cint(\qvec)$ and $\Cext(\qvec)$ 
using the $\Cint(\qvec,\Nk)$ and $\Cext(\qvec,\Nk)$ measured at finite $\Nk$. 
While the $\Nk$-dependent correction term is less crucial for $\Cint(\qvec,\Nk)$,
it is important, as above for $\sOext(\Nk)$, to take advantage of Eq.~(\ref{eq_Cext_Nk}),
especially if $\Cint(\qvec)$ is large.
We focus below on the $\Nk$-extra\-po\-la\-ted correlation functions in real space.

The correlation functions may {\em a priori} also depend explicitly on 
the sampling time $\tsamp$ and the system volume $V$.
One trivial reason for a $V$-dependency is that the linear box length $L$ sets a cut-off.
Fortunately, this only matters for large distances $r \approx L/2$
(and for the corresponding small wavevectors), i.e. this effect becomes irrelevant for large $L$
as one verifies by systematically increasing the box size.
Since $\sOext(\tsamp,V)$ becomes $\tsamp$-independent and $\sOint(\tsamp,V) \propto 1/\sqrt{\tsamp}$ 
for $\tsamp \gg \taubasin$ this naturally suggests 
\begin{equation}
\left. \begin{array}{ll}
\Cext(\rvec,\tsamp,V) \simeq & \Cext(\rvec,V) \\
\Cint(\rvec,\tsamp,V) \propto & 1/\tsamp
\end{array}
\right\}
\mbox{ for } \tsamp \gg \taubasin
\label{eq_corr_tsamp_hyp}
\end{equation}
as discussed theoretically in more detail 
in Appendix~\ref{app_corr_limits} and Appendix~\ref{app_Cint}.
We shall verify numerically in the next subsection whether this holds for our model systems.

\subsection{Examples for lattice spring models}
\label{corr_examples}

\begin{figure}[t]
\centerline{\resizebox{.90\columnwidth}{!}{\includegraphics*{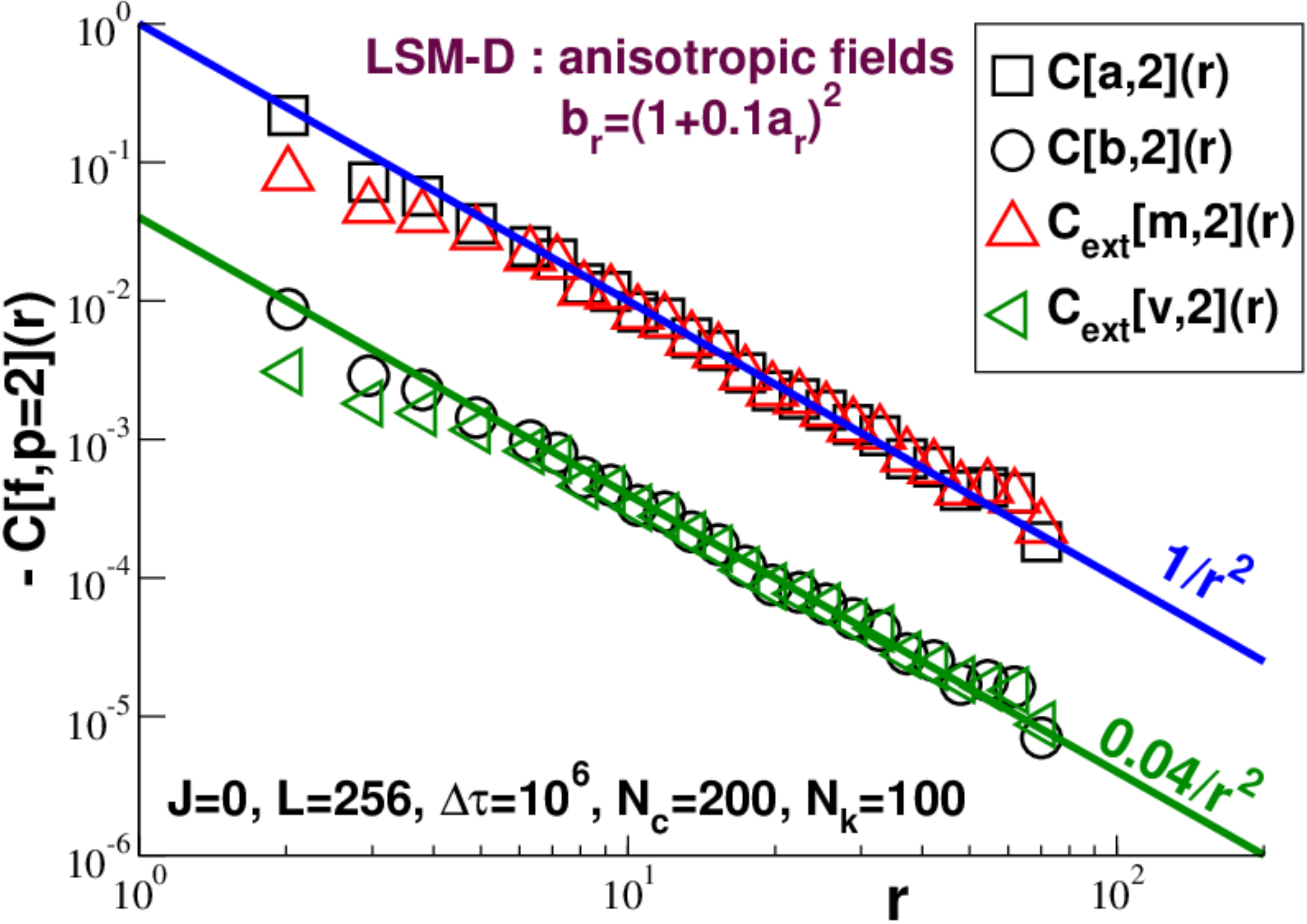}}}
\caption{$-C[f,p=2](r)$ for LSM-D for $J=0$ and $\tsamp \gg \Tnonerg \gg \taubasin$ 
confirming that $C[a](\rvec) \approx \Cext[m](\rvec)$ and $C[b](\rvec) \approx \Cext[v](\rvec)$
as expected from Eq.~(\ref{eq_mrar_vrbr}).
}
\label{fig_Cextr_caseD}
\end{figure}

We present now various (projected) correlation functions $C[f,p](r)$ from our LSM simulations.
We begin in Fig.~\ref{fig_Cextr_caseD} with data from LSM-D obtained 
for $J=0$ and a large sampling time $\tsamp=10^6$.
We remind that LSM-D is defined by Eq.~(\ref{eq_Cfr_modelD}) for the $a$-field 
and by Eq.~(\ref{eq_br_2nd_closure}) for the $b$-field. 
All indicated correlation functions are obtained by anisotropic projection ($p=2$).
Since all spring interactions are switched off ($J=0)$ and since $\tsamp \gg \taubasin$
we have $\mr \approx \ar$ and $\vr \approx \br$.
As expected from Eq.~(\ref{eq_Cfr_modelD}) and Eq.~(\ref{eq_Cb_2nd_closure}), 
Fig.~\ref{fig_Cextr_caseD} confirms 
\begin{eqnarray}
-\Cext[m,p](r) & \approx & -C[a,p](r) \approx  1/r^2 \label{eq_Cmext_modelD} \\
-\Cext[v,p](r) & \approx & -C[b,p](r) \approx (2\lambda)^2/r^2 \label{eq_Cvext_modelD}
\end{eqnarray}
with $p=2$ and $\lambda=0.1$.
Similar results have been found for all model cases with $|J| \ll 1$ and $\tsamp \gg \taubasin$.
Since, moreover, $\tsamp \gg \Tnonerg$ for the presented data, the internal
correlation functions $\Cint[f](\rvec,\tsamp)$ are negligible small and 
$\Ctot[f](\rvec,\tsamp) \approx \Cext[f](\rvec)$ holds (not shown).

\begin{figure}[t]
\centerline{\resizebox{.90\columnwidth}{!}{\includegraphics*{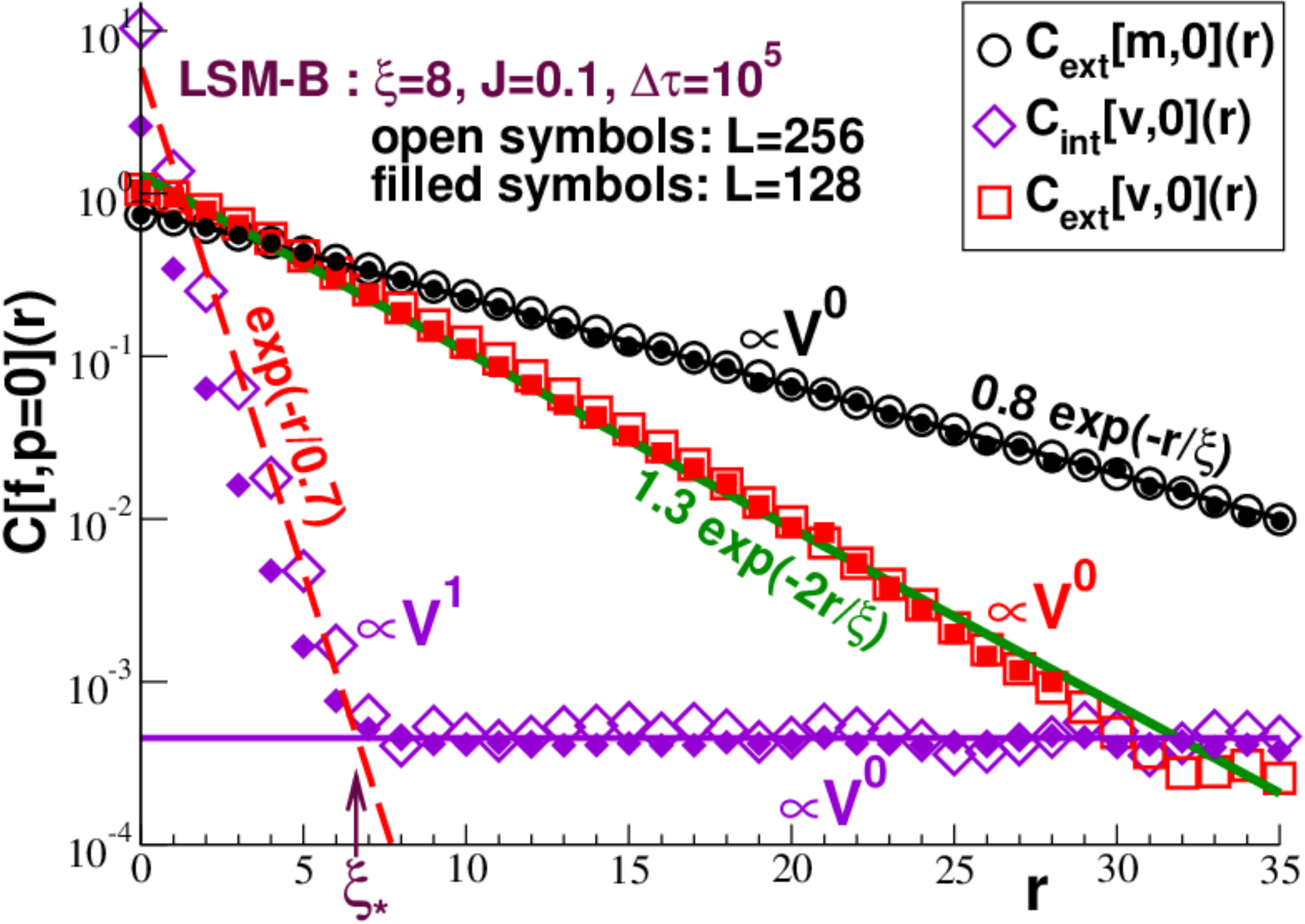}}}
\caption{Isotropic projection $C[f,p=0](r)$ for LSM-B with $\xi=8$, $J=0.1$ and $\tsamp=10^5$.
Open symbols refer to $L=256$, small filled symbols to $L=128$.
As expected $\Cext[m](r)$ and $\Cext[v](r)$ decrease exponentially just as $C[a](r)$ and $C[b](r)$.
$\Cint[v](r)$ decays exponentially (dashed line) for $r \ll \xistar$ 
($\xistar$ being a crossover length)
but becomes constant for large $r$.
}
\label{fig_Cr_caseB}
\end{figure}

All correlation functions presented below in this section 
are isotropically projected. ``$p=0$" is often suppressed for clarity.  
We consider now finite spring interactions and smaller sampling times.
As an example we show in Fig.~\ref{fig_Cr_caseB} 
correlation functions obtained for LSM-B with $\xi=8$, $J=0.1$ and $\tsamp=10^5$.
Data for two system sizes are compared.
$\Cext[m](r)$ and $\Cext[v](r)$ are $V$-independent for all $r \ll L/2$.
The small, finite $J$ only has a minor effect on the prefactors:
As for $J=0$ we observe $\Cext[m](r) \approx C[a](r) \approx \exp(-r/\xi)$ and 
$\Cext[v](r) \approx C[b](r) \approx \exp(-2r/\xi)$. The observed short-range
correlations are consistent with $\gamma=1/2$ (cf. Fig.~\ref{fig_vdvext_V}).

\begin{figure}[t]
\centerline{\resizebox{.90\columnwidth}{!}{\includegraphics*{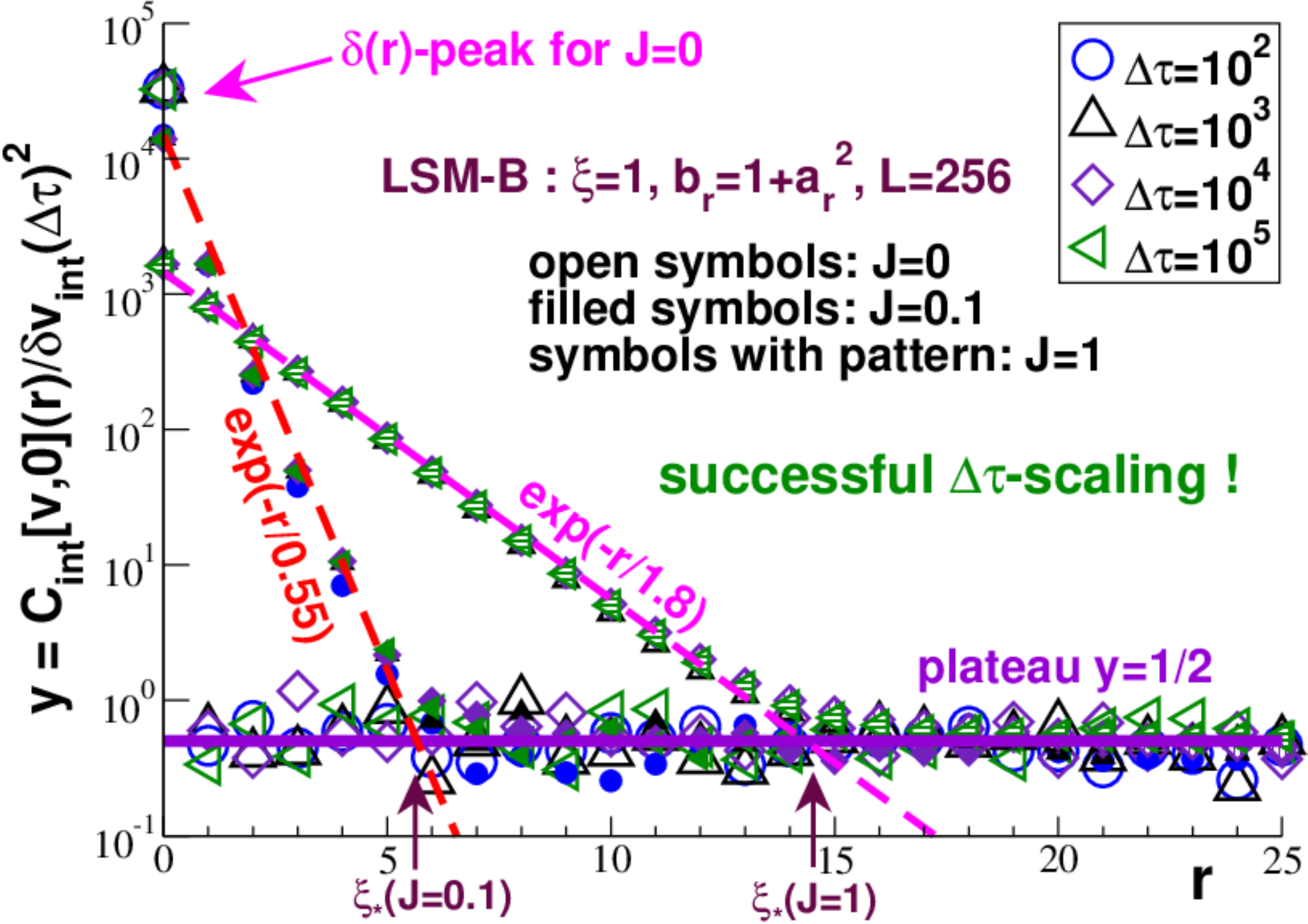}}}
\caption{$y=\Cint[v,0](r)/\dvint(\tsamp)$ {\em versus} $r$ for LSM-B ($\xi=1$, $L=256$) 
with $J=0$ (open symbols), $J=0.1$ (filled symbols) and $J=1$ (symbols with pattern).
Data from a broad $\tsamp$-range scale nicely for each $J$.
Note the huge peak at $r=0$ (arrow) for $J=0$ and the exponential decays (dashed lines) 
with $y(r) \sim V\exp(-r/\xiind)$ for $J >0$ and $r \ll \xistar(J)$.
For large $r$ all data approach $y=1/2$ (bold horizontal line).
}
\label{fig_Cvintr_caseB}
\end{figure}

We turn next to the scaling of the internal correlation function $\Cint[v](r)$.
Focusing on LSM-B this is presented in Fig.~\ref{fig_Cr_caseB} and Fig.~\ref{fig_Cvintr_caseB}.
As we shall see, all our numerical data are consistent with the general scaling
\begin{equation}
\Cint[v](r,\tsamp,V) = [ V (1-\alpha) c(r) + \alpha ] \ \dvint(\tsamp)
\label{eq_Cint_scal}
\end{equation}
with $\alpha = 1/2$ and $c(r)$ being a $\tsamp$-independent function,
depending somewhat on the model (especially on the coupling parameter $J$), 
vanishing for large distances $r$ and being normalized as $V \Eop^{\rvec} c(\rvec)=1$. 
In fact, this scaling is natural for a large class of models as further discussed in Appendix~\ref{app_Cint}.

Let us focus first on the $\tsamp$-dependence of the internal correlation function.
We present in Fig.~\ref{fig_Cvintr_caseB} the rescaled correlation function 
$y = \Cint[v](r)/\dvint(\tsamp)$ as a function of $r$ for LSM-B with $J=0$, $J=0.1$ and $J=1$.
A perfect data collapse is observed for each $J$ confirming thus Eq.~(\ref{eq_Cint_scal}).
Since $\dvint(\tsamp) \propto 1/\tsamp$ for the presented sampling times,
we could have also used as vertical axis $\Cint[v](r,\tsamp) \times \tsamp$ to scale the data.
Importantly, the {\em dimensionless} scaling variable $y$ is more general allowing 
the scaling for all $\tsamp$, i.e. also for $\tsamp \ll \taubasin$.

Turning to the $r$-dependence we note that Eq.~(\ref{eq_Cint_scal}) 
implies that $\Cint[v](r,\tsamp,V)$ should level off to a plateau with $\alpha \dvint(\tsamp) > 0$ 
for sufficiently large $r \gg \xistar$.\footnote{According to Eq.~(\ref{eq_Cint_scal}) and
assuming $c(r)$ to be continuous, the crossover length $\xistar$ may be defined by 
$c(r=\xistar) \approx 1/V$.}
As emphasized by the bold horizontal lines in Fig.~\ref{fig_Cr_caseB} 
and Fig.~\ref{fig_Cvintr_caseB} this is indeed the case. Moreover, the latter figure confirms 
$\alpha=1/2$, i.e. quite generally we have $\Cint[v](r) \to \dvint(\tsamp)/2$ for large $r$.
That $\Cint[v](r)$ becomes a finite constant for large $r$,
albeit the instantaneous $x_{t\rvec}$-field is decorrelated, 
has to do with the definition of the $v_{\rvec}$-field, Eq.~(\ref{eq_Ovr_def}),
as further explained in Appendix~\ref{app_Cint_LSMA}. 
As can be seen from the latter calculation the function $c(r)$ for LSM-A with $J=0$ has 
a jump singularity at $r=0$, Eq.~(\ref{eq_corrCint_F}). That this is also the case for all 
other models with $J=0$ can be seen for LSM-B in Fig.~\ref{fig_Cvintr_caseB} (arrow).
This becomes different if the interaction between the springs is switched on ($J>0$).
As seen in Fig.~\ref{fig_Cr_caseB} and Fig.~\ref{fig_Cvintr_caseB} 
we then observe for $r \ll \xistar$ a continuous exponential decay 
$\Cint[v](r) \propto c(r) \propto \exp(-r/\xiind)$ with a finite induced
correlation length $\xiind$ weakly increasing with $J$.

Moreover, as can be also seen in Fig.~\ref{fig_Cr_caseB},
the internal correlations increase in the first $r$-regime with $V$.
Confirming the $V$-dependence indicated in Eq.~(\ref{eq_Cint_scal}), 
a systematic comparison of a broad range of $L$ reveals that $\Cint[v](r) \propto V$ for small $r$
while it is strictly $V$-independent for large $r$ (not shown).
Due to both contributions the volume average $\Eop^{\rvec}\Cint[v](\rvec)$ is thus 
$V$-in\-de\-pen\-dent consistently with the $V$-independence of $\dvint(\tsamp)$ demonstrated above
(cf. Fig.~\ref{fig_vdv_Dt_J1_caseB} and Sec.~\ref{glob_Vol}).
\footnote{Data collapse for different $V$ and $J>0$ can be achieved (not shown) 
by obtaining first $c(r) = (2y-1)/V$ and by plotting then $c(r)/c(0)$ as a function
of $x = r/\xiind(J,V)$.}

\begin{figure}[t]
\centerline{\resizebox{.90\columnwidth}{!}{\includegraphics*{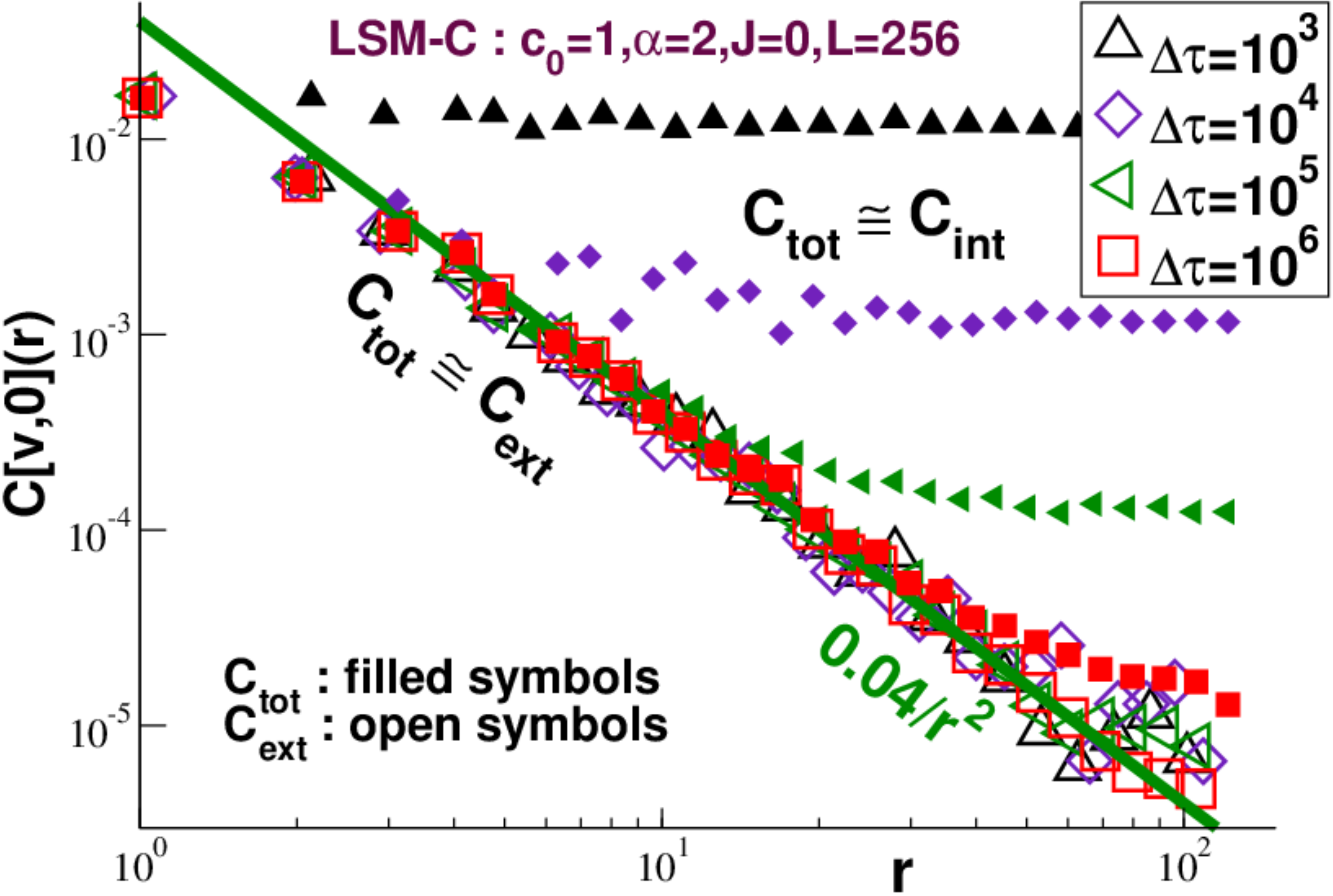}}}
\caption{Isotropic projections $\Ctot[v](r,\tsamp)$ and $\Cext[v](r,\tsamp)$ 
for LSM-C with $\alpha=2$, $J=0$ and $L=256$. Also indicated is the imposed 
asymptotic limit $C[b](r) \approx 0.04/r^2$ (bold solid line) for large $\tsamp$ 
and $V$ for both correlation functions. 
While $\Cext[v](r,\tsamp) \approx C[b](r)$ for $\tsamp \gg \taubasin$,
$\Ctot[v](r,\tsamp)$ is seen to converge much more slowly to the large-$\tsamp$ limit.
}
\label{fig_Cvtotr_caseC}
\end{figure}

The scaling of $\Cext[v](r,\tsamp)$ and $\Ctot[v](r,\tsamp)$
with $\tsamp$ is illustrated in Fig.~\ref{fig_Cvtotr_caseC}. We present here data obtained
for LSM-C with $\alpha=2$, $J=0$ and $L=256$. Importantly, both $\Ctot[v](r,\tsamp)$ and
$\Cext[v](r,\tsamp)$ must approach for sufficiently large $\tsamp$ the (known) asymptotic
limit $C[b](r) \approx 0.04/r^2$ (bold solid line) imposed by construction.
$\Nk=1000$ is used for $\tsamp \le 10^3$ to improve the precision of the
$\Nk$-extrapolation for the external correlation function.
While $\Cext[v](r,\tsamp)$ becomes rapidly $\tsamp$-independent ($\tsamp \gg \taubasin$)
several orders of magnitude larger sampling times are needed for $\Ctot[v](r,\tsamp)$.
This is caused by the internal contribution $\Cint[v](r,\tsamp) \propto 1/\tsamp$
to the total correlation function. This is also responsible for the leveling-off of $\Ctot[v](r,\tsamp)$ 
for large $r \gg \rnonerg(\tsamp)$ with $\rnonerg(\tsamp)$ being 
a crossover distance defined by $\Cint[v](\rnonerg,\tsamp)=\Cext[v](\rnonerg)$.
For the same reasons that $\svtot(\tsamp,V)$ is problematic
for the determination of the system-size exponent $\gamma$, only computing $\Ctot[v](r,\tsamp)$
for {\em one} $\tsamp$ may incorrectly suggest a weak (possibly long-ranged) decay of the correlations.
Only the $r$-regime where the data sets for the largest available $\tsamp$ clearly collapse
can be used. This would be in the presented case less than an order of magnitude.
Similar behavior has been found for all LSM versions. 

\begin{figure}[t]
\centerline{\resizebox{.90\columnwidth}{!}{\includegraphics*{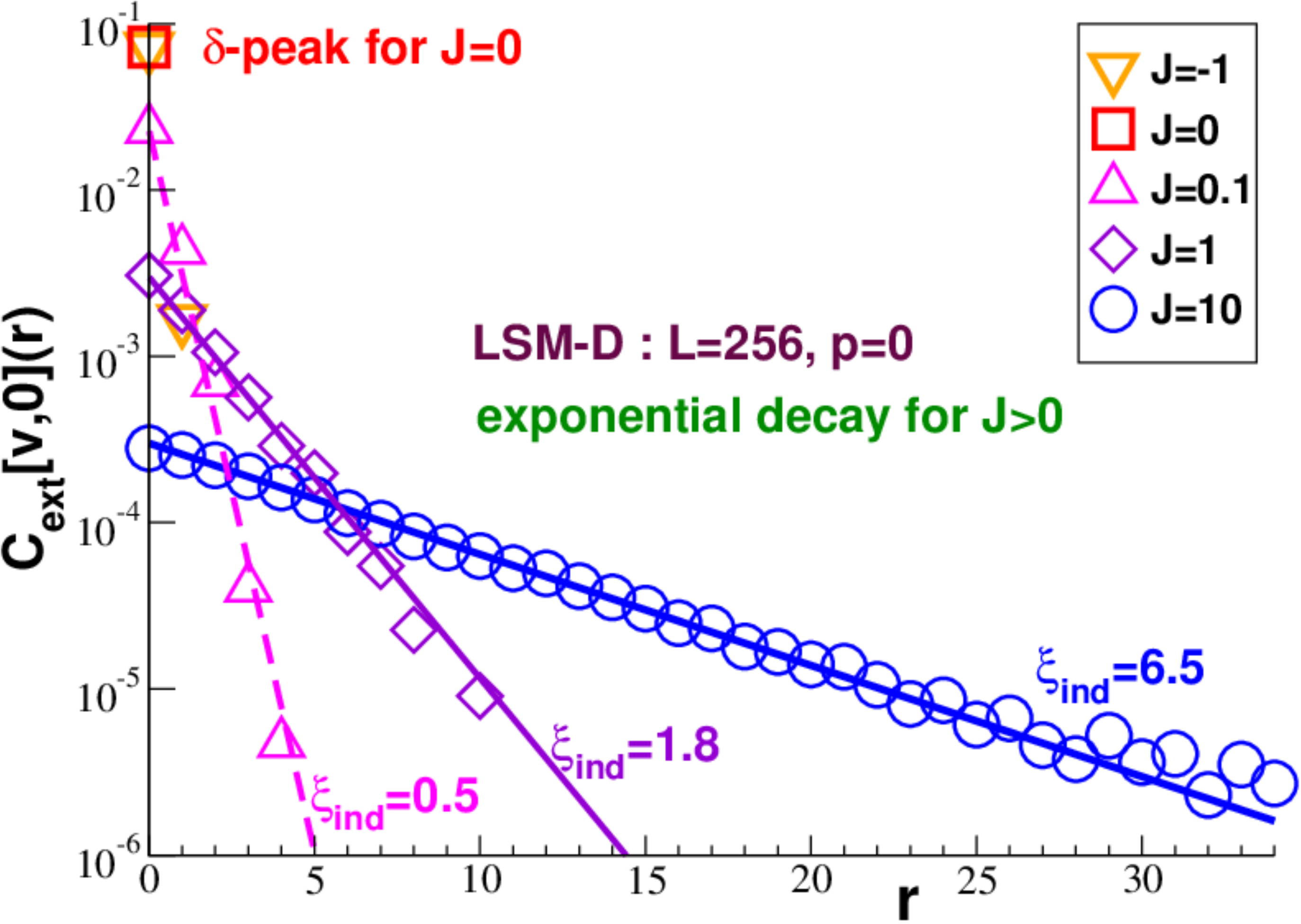}}}
\caption{Isotropic projection $\Cext[v,0](r,\tsamp)$ for LSM-D for different $J$.
A $\delta(r)$-peak is observed for $J=0$ while $\Cext[v,0](r,\tsamp) \propto \exp(-r/\xiind(J))$ for $J>0$.}
\label{fig_Cextr_J_caseD}
\end{figure}

As a last example we come back to LSM-D and present the isotropic ($p=0$)
external correlations $\Cext[v,0](r)$ for different $J$ obtained for $\tsamp \gg \taubasin(J)$. 
This is shown in Fig.~\ref{fig_Cextr_J_caseD} using half-logarithmic coordinates.
As expected from the imposed (quenched) anisotropic $a$- and $b$-fields of the LSM-D (Table~\ref{tab_model}) 
a $\delta(r)$-peak at the origin is observed if all spring interactions are switched off ($J=0$).
At variance with this the external correlation functions decay exponentially for $J > 0$.
Apparently, the corresponding induced correlation length $\xiind(J)$
increases with $J$ but remains finite.
The range of the correlations thus increases but stays {\em short-ranged}
in agreement with the exponent $\gamma=1/2$ seen in Fig.~\ref{fig_dvext_V_caseD}.

\section{Conclusion}
\label{conc}

\subsection{General points made}
\label{conc_generalities}

Extending recent work on stochastic processes in non-ergodic macroscopic systems \cite{spmP1,spmP2,spmP3} 
we have investigated the different types of spatial correlation functions $C(\rvec)$ related to the 
macroscopic variances $\delta \Ocal^2$ of observables $\Ocal[\xbf]$ of time-series $\xbf$.
As reminded in Sec.~\ref{glob} the standard total variance $\dOtot(\tsamp)$
is the sum of an internal variance $\dOint(\tsamp)$ and an external variance $\dOext(\tsamp)$,
cf.~Eq.~(\ref{eq_intro_dOtot}).
While $\sOint(\tsamp) \simeq 1/\sqrt{\tsamp}$ for $\tsamp \gg \taubasin$,
$\sOext(\tsamp,V) \simeq \Snonerg(V)$ becomes constant.
One motivation of this work (cf.~Sec.~\ref{intro_motivation})
was to understand the $V$-dependence of the non-ergodicity parameter $\Snonerg$ in systems 
with (possibly long-ranged) spatial correlations.
The generally important novel key results of this study 
were presented in Sec.~\ref{corr} and Appendix~\ref{app_corr}. As shown there,
assuming $\Ocal[\xbf]$ to be a spatial average of a local field $\Ocal_{\rvec}$,
the three global variances can be written as volume averages of 
the three spatial correlation functions 
$\Ctot(\rvec)$, $\Cint(\rvec)$ and $\Cext(\rvec)$ and, moreover, 
$\Ctot(\rvec) = \Cint(\rvec) + \Cext(\rvec)$ holds.
The $\tsamp$- and $V$-dependences of the global variances can thus be traced back to 
the internal and external correlation functions $\Cint(\rvec)$ and $\Cext(\rvec)$.

\subsection{Specific fields considered}
\label{conc_specificfields}

Focusing on the arithmetic mean $\Ocal[\xbf]=m[\xbf]=\Eop^{t}x_t$ and especially on 
the empirical variance $\Ocal[\xbf]=v[\xbf]=\Vop^{t}x_t$ of time-series (cf. Sec.~\ref{spf})
we illustrated the various general theoretical relations by means of MC simulations of variants of a 
simple lattice spring model (LSM) in two dimensions characterized 
by two quenched and spatial correlated fields (Table~\ref{tab_model}). 
We have especially investigated $\svint(\tsamp,V)$ and $\svext(\tsamp,V)$ and 
the corresponding correlation functions $\Cint[v](\rvec,\tsamp,V)$ and $\Cext[v](\rvec,\tsamp,V)$.
As discussed in Sec.~\ref{corr_examples} and Appendix~\ref{app_Cint}, 
under general assumptions the internal correlation function 
is given by Eq.~(\ref{eq_Cint_scal}),
i.e. it decreases inversely with $\tsamp$ for $\tsamp \gg \taubasin$ and 
becomes constant, $\Cint[v](\rvec,\tsamp,V) \simeq \dvint(\tsamp)/2$, 
for large $r$ (cf. Fig.~\ref{fig_Cvintr_caseB}) 
albeit the primary instantaneous field $x_{t\rvec}$ is decorrelated.
The external correlation function becomes $\tsamp$-inde\-pen\-dent
for $\tsamp \gg \taubasin$ (cf. Fig.~\ref{fig_Cvtotr_caseC}).
The last statement requires a proper $\Nk$-extrapolation by means of Eq.~(\ref{eq_Cext_Nk})
or that the data are sampled using sufficiently large $\tsamp$ and $\Nk$,
i.e. the correction term $\Cint[v](\rvec,\tsamp,V)/\Nk$ in Eq.~(\ref{eq_Cext_Nk}) must be negligible. 
Importantly, $\Cint[v](\rvec,\tsamp) \gg \Cext[v](\rvec)$ for small $\tsamp$, large $V$ and large $r$.
In these limits and due to large crossover effects $\Ctot[v](\rvec,\tsamp)$ 
may deviate from its large-$\tsamp$ limit $\Cext[v](\rvec)$.
Importantly,
this may lead to an overestimation of the range of correlations as shown in Fig.~\ref{fig_Cvtotr_caseC}.
%

\subsection{Outlook}
\label{conc_outlook}

Some of the presented relations and technical features will be used in a 
presentation currently under preparation focusing on shear stresses 
$x_t = \sigma_t$ and associated shear-stress fields $x_{t\rvec}=\sigma_{t\rvec}$ 
\cite{Lemaitre14,Lemaitre15}.
Again focusing on $\tsamp \gg \taubasin$ we investigate $\Cint[f](\rvec)$ and $\Cext[f](\rvec)$ 
for $f=m$ and $f=v$ for a broad range of particle numbers $n \approx V$.
As central results it will be shown that $\Cext[f](\rvec)$ 
is long-ranged for both $f=m$ and $f=v$ decreasing as a power law $1/r^{\alpha}$ with $\alpha \approx d$.
While the scaling of $\Cext[m](\rvec)$ is expected from 
previous simulations \cite{Lemaitre14,Lemaitre15}
and recent theoretical work \cite{Fuchs17,Fuchs18,Fuchs19,lyuda18} the long-range decay of
$\Cext[v](\rvec)$ is non-trivial and, as we shall discuss, indicates that the
corresponding local elastic field is also long-ranged.
 
\section*{Author contribution statement}
JPW designed and wrote the project benefiting from contributions of all authors.

\section*{Acknowledgments}
We acknowledge computational re\-sources 
from the HPC cluster of the University of Strasbourg.

\section*{Data availability statement}
It was not possible to store all the $\Nc \times \Nk \times \Nt \times L^2$ primary fields
$x_{ckt\qvec}$ which were immediately deleted after having been analyzed. 
Tables of the global averages have been kept, however, 
for a broad range of $\tsamp$ and $V$ and different LSM variants. 
These data sets are available from the corresponding author on reasonable request.

\appendix

\section{Spatial correlations of periodic microcells}
\label{app_corr}

\subsection{Some useful general relations}
\label{app_corr_useful}

Let us begin by stating several useful general relations for spatial correlation functions
of $d$-di\-men\-sion\-al, real, discrete and periodic fields.
As defined by Eq.~(\ref{eq_Khat_def}) or Eq.~(\ref{eq_WKT}) we consider 
the instantaneous correlation function $\Khat[\yr](\rvec)$ of a field $\yr$
of volume average $y \equiv \Eop^{\rvec} y_{\rvec}$. Obviously,
\begin{equation}
\Khat[\yr-y](\rvec) = \Khat[\yr](\rvec) - y^2.
\label{eq_K_y_shift}
\end{equation}
Let us assume that the field $y_{l\rvec}$ additionally depends on an index $l$. 
We use below the averages $y_l=\Eop^{\rvec} y_{l\rvec}$, $y_{\rvec}=\Eop^l y_{l\rvec}$ and 
$y=\Eop^l y_l = \Eop^{\rvec} y_{\rvec}$. 
Rewriting Eq.~(\ref{eq_K_y_shift}) and summing over $l$ gives 
\begin{equation}
\Eop^l\Khat[y_{l\rvec}-y_l](\rvec) = \Eop^l \Khat[y_{l\rvec}](\rvec) - \Eop^ly_l^2.
\label{eq_K_yl_shift}
\end{equation}
Also it is seen by expansion using Eq.~(\ref{eq_Khat_def}) that
\begin{equation}
\Eop^lK[y_{l\rvec}-\lambda y_{\rvec}](\rvec)
= \Eop^lK[y_{l\rvec}](\rvec) - \lambda (2-\lambda) K[y_{\rvec}](\rvec)
\label{eq_K_yr_shift}
\end{equation}
for any real constant $\lambda$.
 
We remind that $\Vop^l y_l = \Eop^l y_l^2 - y^2 = \Eop^l (y_l-y)^2$.
Using again Eq.~(\ref{eq_K_y_shift}) and the periodicity of the grid 
the variance $\Vop^l y_l$ may be written as the volume average 
\begin{eqnarray}
\Vop^l y_l & = &  \Eop^{\rvec} C(\rvec) \mbox{ with } \label{eq_Cr_A} \\
C(\rvec) & \equiv & \Eop^l K[y_{l\rvec}-y](\rvec) \equiv \Eop^l K[y_{l\rvec}](\rvec) - y^2 
\label{eq_Cr_B}
\end{eqnarray}
being the $l$-averaged correlation function in real space.
By comparing Eq.~(\ref{eq_K_yl_shift}) and Eq.~(\ref{eq_Cr_B}) we may also write 
\begin{equation}
C(\rvec) = \Eop^l K[y_{l\rvec}-y_l](\rvec) + \Vop^l y_l \label{eq_Cr_C}
\end{equation}
which using Eq.~(\ref{eq_Cr_A}) implies $\Eop^{\rvec}\Eop^l K[y_{l\rvec}-y_l](\rvec) =0$.
It is useful to state two important limits for $C(\rvec)$:
{\em (i)} At the origin we have
\begin{equation}
C(\rvec=\bfzero) = \Vop^{l\rvec'} y_{l\rvec'} = 
\Eop^l \Vop^{\rvec'} y_{l\rvec'} + \Vop^l \Eop^{\rvec'} y_{l\rvec'}
\label{eq_Crzero}
\end{equation}
and {\em (ii)} $C(\rvec)$ exactly vanishes if and only if 
\begin{equation}
\Eop^l \Eop^{\rvec'} y_{\rvec'+\rvec} y_{\rvec'} =
\Eop^l \Eop^{\rvec'} y_{\rvec'+\rvec} \times \Eop^l \Eop^{\rvec'} y_{\rvec'} = y^2
\label{eq_Crvanishes}
\end{equation}
as it happens for most (albeit not all) fields for sufficiently large $r=|\rvec|$.
See Appendix~\ref{app_Cint_LSMA} for an exception relevant for the present study.
 
Moreover, with $z=\Eop^{\rvec} z_{\rvec}$ being an $l$-independent quantity
it follows from Eq.~(\ref{eq_Cr_A}) that 
\begin{equation}
\Vop^l (y_l-z) = \Eop^{\rvec} \ \Eop^l K[(y_{l\rvec}-z_{\rvec})-(y-z)](\rvec).
\label{eq_Cr_z_shift}
\end{equation}
Since $\Vop^l y_l = \Vop^l (y_l - z)$ this implies quite generally that
\begin{equation}
\Eop^{\rvec} \Eop^l K[\delta y_{l\rvec}](\rvec) = \Eop^{\rvec} \Eop^l K[\delta y_{l\rvec}-\delta z_{\rvec}](\rvec)
\label{eq_Cr_z_shift_Er}
\end{equation}
with $\delta y_{l\rvec}=y_{l\rvec}-y$ and $\delta z_{\rvec}=z_{\rvec}-z$,
i.e. the correlation function of a field $y_{l\rvec}$ can be shifted by an 
$l$-independent field $z_{\rvec}$ {\em without} changing the $l$-averaged volume average.

\subsection{Derivation of correlation functions}
\label{app_corr_inexto}

Using these general relations it is readily seen that the correlation functions defined as
\begin{eqnarray}
\hspace*{-0.5cm}\Ctot(\rvec) & \equiv & \Eop^l K[\Ocal_{l\rvec}-\Ocal](\rvec)
= \Eop^l K[\Ocal_{l\rvec}](\rvec) - \Ocal^2
\label{eq_Ctotr_def} \\
\hspace*{-0.5cm}\Cext(\rvec) & \equiv & \Eop^c K[\Ocr-\Ocal](\rvec) 
= \Eop^c K[\Ocr](\rvec) - \Ocal^2
\label{eq_Cextr_def} \\
\hspace*{-0.5cm}\Cint(\rvec) & \equiv & \Eop^c \Eop^k K[\Ockr-\Ocr](\rvec)
\label{eq_Cintr_def} 
\end{eqnarray}
are consistent with Eqs.~(\ref{eq_intro_Ctotr2dOtot}-\ref{eq_intro_Cinexto}).
The index $l$ runs again over all independent configurations $c$ and 
all time-series $k$ for each $c$ and the expectation value $\Ocal$
is defined in Eq.~(\ref{eq_EcEk_commute}).
The corresponding equations in reciprocal space are given in Sec.~\ref{corr_inexto},
Eqs.~(\ref{eq_Ctotq_def}-\ref{eq_Cextq_def}).
That Eq.~(\ref{eq_Ctotr_def}) is consistent with $\dOtot = \Eop^{\rvec} \Ctot(\rvec)$
and Eq.~(\ref{eq_Cextr_def}) with $\dOext = \Eop^{\rvec} \Cext(\rvec)$
is directly implied by Eq.~(\ref{eq_Cr_A}) and Eq.~(\ref{eq_Cr_B}).
To show that Eq.~(\ref{eq_Cintr_def}) is consistent with $\dOint = \Eop^{\rvec} \Cint(\rvec)$
and that all three correlation functions sum up according to Eq.~(\ref{eq_intro_Cinexto}) 
let us first note that due to Eq.~(\ref{eq_K_yr_shift}) for $\lambda=1$
the internal correlation function may be rewritten as
\begin{equation}
\Cint(\rvec) = \Eop^c \left\{\Eop^k K[\Ockr](\rvec) - K[\Ocr](\rvec)\right\}.
\label{eq_Cintr_def_B}
\end{equation}
Using Eq.~(\ref{eq_Ctotr_def}) and Eq.~(\ref{eq_Cextr_def}) this implies
$\Cint(\rvec) = \Ctot(\rvec) - \Cext(\rvec)$ in agreement with 
the key relation Eq.~(\ref{eq_intro_Cinexto}) stated in the Introduction.
In turn we thus have 
\begin{eqnarray}
\Eop^{\rvec} \Cint(\rvec) & = & \Eop^{\rvec} \left( \Ctot(\rvec) - \Cext(\rvec)\right) \nonumber \\
& = & \dOtot - \dOext = \dOint \label{eq_dOintCint}
\end{eqnarray}
where we have used Eq.~(\ref{eq_intro_dOtot}) in the last step.

Please note that due to Eq.~(\ref{eq_Cr_z_shift_Er}) $\dOint = \Eop^{\rvec} \Cint(\rvec)$ 
would also be solved by the more general internal correlation function 
\begin{equation}
\Cint(\rvec,\lambda) \equiv \Eop^c \Eop^k K[(\Ockr-\Ocal_c)-\lambda (\Ocr-\Ocal_c)](\rvec)
\label{eq_Cint_A}
\end{equation}
which reduces to Eq.~(\ref{eq_Cintr_def}) for $\lambda=1$,
since it is possible to shift $\Ockr-\Ocal_c$ with the $k$-independent field 
$\lambda (\Ocal_{c\rvec}-\Ocal_c)$ without changing the $k$-averaged volume average.
The trouble with such alternative definitions is that Eq.~(\ref{eq_intro_Cinexto})
does not hold anymore in general, e.g., it can be shown that Eq.~(\ref{eq_Cint_A})
leads to
\begin{equation}
\Ctot(\rvec) = \Cint(\rvec,\lambda) + \lambda (2-\lambda) \Cext(\rvec) + (\lambda-1)^2 \dOext.
\label{eq_Cint_D}
\end{equation}
Due to the last term and since $\dOext > 0$ for non-ergodic systems all three correlation functions 
may in principle only vanish for the same $\rvec$ for $\lambda =1$ and for exactly this limit 
Eq.~(\ref{eq_Cint_D}) reduces to Eq.~(\ref{eq_intro_Cinexto}).
We therefore set $\lambda=1$.

\subsection{Important limits}
\label{app_corr_limits}

We have omitted for clarity in the preceding subsection
all possible dependences on $\Nc$, $\Nk$, $\tsamp$ and $V$.
However, it is assumed below that $\Nc$ and $\Nk$ are arbitrarily large,
i.e. all properties are $\Nc$- and $\Nk$-independent.
Moreover, we focus on the limit $\tsamp \gg \taubasin$,
i.e. both $\Ocal_{c\rvec} = \Eop^k \Ocal_{ck\rvec}$ and 
its average $\Ocal = \Eop^c \Eop^{\rvec} \Ocal_{c\rvec}$
are $\tsamp$-independent to leading order. Due to Eq.~(\ref{eq_Cextr_def}) the same holds
for the external correlation function, i.e.
\begin{equation}
\Cext(\rvec,\tsamp,V) \simeq \Cext(\rvec,V)
\mbox{ for } \tsamp \gg \taubasin.
\label{eq_corrlimit_Cext}
\end{equation}
The indicated $V$-dependence drops out if $\Ocal_{c\rvec}$ is $V$-in\-de\-pen\-dent 
as in all the models of this work.

The internal correlation function, Eq.~(\ref{eq_Cintr_def}),
characterizes the correlations of the difference $\Ocal_{ck\rvec}(\tsamp) - \Ocal_{c\rvec}$.
While $\Ocal_{ck\rvec}(\tsamp)$ depends in general not only on $k$ but also on $\tsamp$,
{\em both} dependences drop out for $\tsamp \to \infty$.
Hence, $\Ocal_{ck\rvec}(\tsamp) \to \Ocal_{c\rvec}$ and in turn 
\begin{equation}
\lim_{\tsamp \to \infty} \Cint(\rvec,\tsamp,V) = 0.
\label{eq_Cint_tsamp_infty}
\end{equation}
To obtain the internal correlation function for {\em finite} $\tsamp \gg \taubasin$ 
it should be remembered that $\Ocal_{ck\rvec}(\tsamp)$ is a time-averaged moment
over $\Nt=\tsamp/\tincr$ data entries from one stored time-series. 
The internal correlation function can thus be written as an average
\begin{equation}
\Cint(\rvec=\rvec_2-\rvec_1,\tsamp,V) =
\Eop^c \Eop^{\rvec_1} \left( \Eop^{t_1} \Eop^{t_2} \Eop^k \ldots\right) 
\label{eq_Cint_T}
\end{equation} 
over entries measured at discrete times $t_1$ and $t_2$.
A specific example is worked out in Appendix~\ref{app_Cint_LSMA}.
If one assumes for simplicity that $\tincr \gg \taubasin$ 
only contributions with $t_1=t_2$ can contribute.
Using also that the time-average $\Eop^t$ is normalized by $\Nt \propto \tsamp$,
this shows that quite generally the internal correlation function must decay 
to leading order for all $\rvec$ as
\begin{equation}
\Cint(\rvec,\tsamp,V) \propto \frac{1}{\tsamp} \mbox{ for } \tsamp \gg \taubasin
\label{eq_corrlimit_Cint}
\end{equation}
as expected from $\dOint(\tsamp) = \Eop^{\rvec} \Cint(\rvec) \propto 1/\tsamp$.

\section{Scaling of $\Cint[v](\rvec,\tsamp,V)$}
\label{app_Cint}

\subsection{Predictions for LSM-A}
\label{app_Cint_LSMA}

\begin{table*}[t]
\begin{center}
\begin{tabular}{|c||l|l|l|l|}
\hline
case & $\Eop^k A_1A_2$    & $\Eop^k A_1B_2$  & $\Eop^k A_2B_1$  & $\Eop^k B_1B_2$  \\ \hline
1. $\rvec_1=\rvec_2=\rvec_3=\rvec_4$ & 
$2b_{c\rvec_1}^2/\Nt$     & $3b_{c\rvec_1}^2/\Nt^2$   & $3b_{c\rvec_1}^2/\Nt^2$    &  $b_{c\rvec_1}^2/\Nt^3$ \\
2. $\rvec_1=\rvec_2 \ne \rvec_3=\rvec_4$ & 
$b_{c\rvec_1} (\Sop^{\rvec_3\ne\rvec_1}b_{c\rvec_3})/\Nt$
& $b_{c\rvec_1}^2 (V-1)/\Nt^2$   &  $b_{c\rvec_1}^2 (V-1)/\Nt^2$   &  $b_{c\rvec_1}^2 (V-1)/\Nt^2$ \\
3. $\rvec_1=\rvec_3 \ne \rvec_2=\rvec_4$ & 
0    & 0   &  0   &  $b_{c\rvec_1}b_{c\rvec_2}/\Nt^2$ \\
4. $\rvec_1=\rvec_4 \ne \rvec_2=\rvec_3$ &
$b_{c\rvec_1}b_{c\rvec_2}/\Nt$  & $b_{c\rvec_1}b_{c\rvec_2}/\Nt^2$   &  $b_{c\rvec_1}b_{c\rvec_2}/\Nt^2$  & $b_{c\rvec_1}b_{c\rvec_2}/\Nt^2$  \\ \hline
\end{tabular}
\caption[]{$k$-averages $\Eop^k A_1A_2$, $\Eop^k A_1B_2$, $\Eop^k A_2B_1$ and $\Eop^k B_1B_2$ for LSM-A.
The different relevant cases for $\rvec_1$, $\rvec_2$, $\rvec_3$, $\rvec_4$ are indicated in the first column.
Note that $r = |\rvec|$ with $\rvec =\rvec_2-\rvec_1$.
The first two cases indicate contributions for $r=0$,
the last two cases contributions for $r>0$.
$\Eop^k A_1B_2$ and $\Eop^k A_2B_1$ are identical by symmetry.
The leading contributions of order $\Ocal(\tsamp^{-1})$ are due to $\Eop^k A_1A_2$ (second column).
The second case ($\rvec_1=\rvec_2 \ne \rvec_3=\rvec_4$) yields contributions proportional to the system size.}
\label{app_tab}
\end{center}
\end{table*}

As noted in Appendix~\ref{app_corr_useful}, all correlation functions $C[f](\rvec)$
discussed in the present work must vanish {\em if} Eq.~(\ref{eq_Crvanishes}) holds, 
i.e. if two typical points of the field $f$ at a respective distance $\rvec$ are uncorrelated. 
Here we draw attention to the fact that although the primary instantaneous field $x_{t\rvec}$ 
may be uncorrelated this may not be the case for the field $\Ocal_{\rvec}$ associated to the 
time-averaged functional $\Ocal[\xbf]$ of the time-series $\xbf$.\footnote{As
seen using Eq.~(\ref{eq_corrCint_xcorr_D}) this is, however, the case for $m_{\rvec} = \Eop^t x_{t\rvec}$ 
for decorrelated primary instantaneous fields $x_{t\rvec}$.}
As we shall see, this matters specifically for the covariance field, Eq.~(\ref{eq_Ovr_def}),
\begin{eqnarray}
v_{ck\rvec} & = & V \left\{ \Eop^t x_{ckt\rvec}x_{ckt} - x_{ck\rvec} x_{ck} \right\} \nonumber \\
 & = & \Sop^{\rvec'} \left\{ 
\Eop^t x_{ckt\rvec} x_{ckt\rvec'} - \Eop^t \Eop^{t'} x_{ckt\rvec} x_{ckt'\rvec'}\right\}
\label{eq_corrCint_vckr}
\end{eqnarray}
restated for convenience omitting the irrelevant prefactor $\beta$
and introducing the sum $\Sop^{\rvec'} \equiv V\Eop^{\rvec'}$.
In the second step we have made explicit the crucial contributions with $\rvec'\ne \rvec$ 
to the averages $x_{ckt}$ and $x_{ck}$.

Without restricting the generality of the argument let us focus on the
model LSM-A with $J=0$, i.e. the $x_{ckt\rvec}$ of different microcells $\rvec$ are uncorrelated.
Ultimately, we want to expand $v_{ck\rvec}$ for large $\tsamp$.
Reminding Eq.~(\ref{eq_mrar_vrbr}) it is thus useful that the $x_{ckt\rvec}$ in Eq.~(\ref{eq_corrCint_vckr}) 
can be replaced by $\delta x_{ckt\rvec} \equiv x_{ckt\rvec}-a_{c\rvec}$ without changing $v_{ck\rvec}$.
Moreover, let us assume that the data is sampled with large time increments $\tincr \gg\taubasin$,
i.e. different times $t'$ and $t''$ can be considered to be uncorrelated. 
This implies that
\begin{equation}
\Eop^k \delta x_{ckt'\rvec'}^{p'} \delta x_{ckt''\rvec''}^{p''}
= \Eop^k \delta x_{ckt'\rvec'}^{p'} \Eop^k \delta x_{ckt''\rvec''}^{p''}
\label{eq_corrCint_xcorr_A}
\end{equation}
whenever $t'\ne t''$ or $\rvec'\ne\rvec''$ ($p'$ and $p''$ being integers) 
and, moreover, the following useful relations hold:
\begin{eqnarray}
	\Eop^{t}\Eop^{t'} \delta_{tt'} & = & 1/\Nt, \label{eq_corrCint_xcorr_B} \\
\Eop^k \delta x_{ckt\rvec} & = & 0, \label{eq_corrCint_xcorr_C} \\
\Eop^k \delta x_{ckt\rvec} \delta x_{ckt'\rvec'} & = & b_{c\rvec} \ \delta_{\rvec\rvec'} \delta_{tt'},
\label{eq_corrCint_xcorr_D}\\
\Eop^k \delta x_{ckt\rvec}^2 \delta x_{ckt\rvec'}^2 & = & 
3 b_{c\rvec}^2 \ \delta_{\rvec\rvec'}  + b_{c\rvec}b_{c\rvec'}(1-\delta_{\rvec\rvec'}). \label{eq_corrCint_xcorr_E} 
\end{eqnarray}
We have used above that the lattice springs are Gaussian variables, Eq.~(\ref{eq_Erspring}),
i.e.  for each independent spring of length $x_{c\rvec}$
(dropping the indices $k$ and $t$) we have 
$\la \delta x_{c\rvec} \ra = 0$, $\la \delta x_{c\rvec}^2 \ra = b_{c\rvec}$ and 
$\la \delta x_{c\rvec}^4 \ra = 3 b_{c\rvec}^2$
with $\la \ldots \ra$ denoting the thermal average within each basin $c$.
Using Eqs.~(\ref{eq_corrCint_xcorr_B}-\ref{eq_corrCint_xcorr_D}) we obtain, e.g., the $k$-average
\begin{equation}
v_{c\rvec} = \Eop^{k} v_{ck\rvec} = b_{c\rvec} \left(1-1/\Nt\right)
\label{eq_corrCint_vcr}
\end{equation}
and, hence, the total average $v=\Eop^c\Eop^{\rvec} v_{c\rvec} = b (1-1/\Nt)$
with $b= \Eop^c\Eop^{\rvec} b_{c\rvec}$.

To compute the internal correlation function we write 
\begin{eqnarray}
\Cint[v](\rvec) & = & \Eop^c C_c(\rvec) \mbox{ with } \label{eq_corrCint_A} \\
C_c(\rvec) & = & \Eop^k K[(v_{ck\rvec}-b_{c\rvec})-(v_{c\rvec}-b_{c\rvec})](\rvec)  \label{eq_corrCint_B} \\
& = & \Eop^k K[v_{ck\rvec}-b_{c\rvec}](\rvec) - K[v_{c\rvec}-b_{c\rvec}](\rvec) \label{eq_corrCint_C} 
\end{eqnarray}
using Eq.~(\ref{eq_Cintr_def_B}) in the last step. Due to Eq.~(\ref{eq_corrCint_vcr})
it is clear that the second term in Eq.~(\ref{eq_corrCint_C}) is of order $\Ocal(\tsamp^{-2})$
We can focus on the first term which we rewrite as volume average 
$\Eop^{\rvec_1} \tilde{C}_c(\rvec_1,\rvec_2=\rvec_1+\rvec)$ with
\begin{eqnarray}
\tilde{C}_{c\rvec_1\rvec_2}(\rvec_1,\rvec_2) & \equiv & \Eop^k (v_{ck\rvec_1}-b_{c\rvec_1}) (v_{ck\rvec_2}-b_{c\rvec_2})
\label{eq_corrCint_D} \\
& = & \Eop^k (A_1 - B_1) (A_2 -B_2) \label{eq_orrCint_E} 
\end{eqnarray}
where Eq.~(\ref{eq_corrCint_vckr}) is used for the definition of the terms
\begin{eqnarray}
A_1 & = & \Sop^{\rvec_3} \Eop^{t_1} 
\left(\delta x_{ckt_1\rvec_1} \delta x_{ckt_1\rvec_3} - b_{c\rvec_1} \delta_{\rvec_1\rvec_3}\right),
\label{eq_corrCint_A1}\\
A_2 & = & \Sop^{\rvec_4} \Eop^{t_2}
\left(\delta x_{ckt_2\rvec_2} \delta x_{ckt_2\rvec_4} - b_{c\rvec_2} \delta_{\rvec_2\rvec_4}\right),
\label{eq_corrCint_A2}\\
B_1 & = & \Sop^{\rvec_3} \Eop^{t_1} \Eop^{t_3} \delta x_{ckt_1\rvec_1} \delta x_{ckt_3\rvec_3} \ \mbox{and}
\label{eq_corrCint_B1}\\
B_2 & = & \Sop^{\rvec_4} \Eop^{t_2} \Eop^{t_4} \delta x_{ckt_2\rvec_2} \delta x_{ckt_4\rvec_4}.
\label{eq_corrCint_B2}
\end{eqnarray}
We expand the different contributions and $k$-average using the identities 
Eqs.~(\ref{eq_corrCint_xcorr_A}-\ref{eq_corrCint_xcorr_E}).
As summarized in Table~\ref{app_tab} it is helpful to distinguish
four cases for $\rvec_1$, $\rvec_2$, $\rvec_3$ and $\rvec_4$.
Most contributions are of order $\Ocal(\tsamp^{-2})$ 
and only three contributions of $\Eop^kA_1A_2$ (second column) do matter.
A central result is that due to the last case ($\rvec_1=\rvec_4 \ne \rvec_2=\rvec_3$), 
the internal correlation function must remain finite for $r > 0$.
Note also that all terms for the second case ($\rvec_1=\rvec_2 \ne \rvec_3=\rvec_4$) 
increase linearly with $V$ due to the sum 
$\Sop^{\rvec_3}\Sop^{\rvec_4} \delta_{\rvec_3\rvec_4} (1-\delta_{\rvec_1\rvec_3})$.
In fact, using $b_c=\Eop^{\rvec_3}b_{c\rvec_3}$, the indicated term for $\Eop^kA_1A_2$ can be rewritten as
\begin{equation}
b_{c\rvec_1} (\Sop^{\rvec_3\ne\rvec_1}b_{c\rvec_3})/\Nt = b_{c\rvec_1} b_c V/\Nt - b_{c\rvec_1}^2/\Nt.
\label{eq_corrCint_A1A2case2}
\end{equation}
We finally average over $\rvec_1$ and $c$ using that the $b_{c\rvec'}$ are decorrelated
for different $\rvec'$. Summarizing all terms we obtain to leading order
\begin{equation}
\Cint[v](\rvec) \simeq \frac{b^2}{\Nt} \times \left\{ 
\begin{array}{ll}
V+1 & \mbox{ for } r = 0,\\
1   & \mbox{ for } r > 0. 
\end{array}
\right.
\label{eq_corrCint_F}
\end{equation}
As a consequence $\dvint(\tsamp,V) = \Eop^{\rvec} \Cint[v](\rvec) = 2b^2/\Nt \propto V^0/\tsamp$,
as expected.

\subsection{Scaling for general Gaussian fields}
\label{app_Cint_gauss}

It is clear that the above result can be generalized to other models
with short-range correlations and general $\tincr$ including $\tincr \ll \taubasin$.
This merely requires a renormalization of space and time.
Especially this suggests to replace $1/\Nt$ by $\dvint(\tsamp)$.
It is in this context of relevance that the above result Eq.~(\ref{eq_corrCint_F}) can be recast as
\begin{equation}
\Cint[v](\rvec) \simeq \left\{V (1-\alpha) c(\rvec)+\alpha\right\} \dvint(\tsamp)
\label{eq_Cint_gauss_A}
\end{equation}
with $\Sop^{\rvec} c(\rvec) = 1$, $c(\rvec) \to 0$ for large $r$ (with $r \ne 0$ for LSM-A) and $\alpha=1/2$.
Note that $\Eop^{\rvec} \Cint[v](\rvec) = \dvint(\tsamp)$ holds for all coefficients $\alpha$.
As discussed in Sec.~\ref{corr_examples}, the numerical results of all our LSM variants are consistent 
with this generalization of the direct calculation for the simple LSM-A model.
(As far as we can tell this even holds reasonably for systems with long-range correlations.)
There is in fact a general reason for expecting Eq.~(\ref{eq_Cint_gauss_A}) to hold for many models:
For a given $c$ the $x_{t\rvec}$-fluctuations are often nearly Gaussian.
[For the LSM variants the joint distributions of the $x_{t\rvec}$ are in fact exactly Gaussian
since the total energy is quadratic in $x_{t\rvec}$, Eq.~(\ref{eq_Erspring}).]
This allows for a theoretical treatment of $\Cint[v](\rvec)$ based on the cumulant
formalism (``Wick's theorem") similar to the calculation which leads to Eq.~(\ref{eq_svint_tsamp})
for the global variance $\dvint(\tsamp)$ \cite{lyuda18,spmP1}. 
It is thus possible to show that Eq.~(\ref{eq_Cint_gauss_A}) must hold 
for general fluctuating Gaussian fields with $\alpha=1/2$.
This calculation is beyond of the scope of the present work.


\end{document}